\documentclass{report}
\usepackage[utf8]{inputenc}
\usepackage[margin=1in,letterpaper]{geometry}
\usepackage{tikz}
\usepackage{tocloft}
\usepackage{sectsty}
\usepackage{titlesec}
\usepackage{setspace}
\usepackage[section]{placeins}
\usepackage{mdframed}
\usepackage{cite}
\usepackage{rotating}
\usepackage[font=small]{caption}
\usepackage[nottoc]{tocbibind}
\sectionfont{\Large}
\usepackage[none]{hyphenat}
\usepackage{url}
\newlength\mylength
\renewcommand\cftchappresnum{\chaptername~}
\renewcommand\cftchapaftersnum{:}
\newcommand\frontmatter{%
    \cleardoublepage
  \pagenumbering{roman}}

\newcommand\mainmatter{%
    \cleardoublepage
  \pagenumbering{arabic}}

\newcommand\backmatter{%
  \if@openright
    \cleardoublepage
  \else
    \clearpage
  \fi
   }
  \titleformat{\chapter}[display]
  {\LARGE \bfseries}
  {\chaptertitlename\ \thechapter}{1em}{\LARGE}
\begin{document}
\sloppy
\frontmatter
\begin{tikzpicture}[remember picture,overlay]
\node [yshift=-2in] at (current page.north)
[font=\Large, anchor=north, text centered]
{
\MakeUppercase{Improving Gene Trees without more data}
};
\end{tikzpicture}
\begin{tikzpicture}[remember picture,overlay]
\node [yshift=-3.5in] at (current page.north)
[font=\Large, anchor=north, text centered,align=center]
{
BY\\\\ASHU GUPTA
};
\end{tikzpicture}
\begin{tikzpicture}[remember picture,overlay]
\node [yshift=-5.5in] at (current page.north)
[font=\Large, anchor=north, text centered,align=center]
{
THESIS\\\\Submitted in partial fulfillment of the requirements\\for the degree of Master of Science in Computer Science\\in the Graduate College of the\\University of Illinois at Urbana-Champaign, 2016
};
\end{tikzpicture}
\begin{tikzpicture}[remember picture,overlay]
\node [yshift=-7.5in] at (current page.north)
[font=\Large, anchor=north, text centered,align=center]
{
Urbana, Illinois
};
\end{tikzpicture}
\begin{tikzpicture}[remember picture,overlay]
\node [yshift=-8.0in, xshift=-3.25in] at (current page.north)
[font=\Large, anchor=north west,align=left]
{
Adviser:\\\\\hspace{0.5in}Professor Tandy Warnow
};
\end{tikzpicture}
\thispagestyle{empty}
\newpage
\section*{ABSTRACT}
\begin{doublespace}
  Species tree and gene tree estimation from sequence data are two steps in many biological analyses. Computational challenges and limited amount of data often make estimating highly accurate phylogenetic trees a difficult task. Moreover, gene alignments used to estimate trees on individual loci often have low phylogenetic signal (e.g., short alignment length), resulting in poorly estimated gene trees. Species tree estimation on the other hand is challenged by individual loci having different evolutionary histories caused by a biological phenomenon known as incomplete lineage sorting (ILS). In the presence of ILS, summary methods like MP-EST, ASTRAL2, and ASTRID are often used to estimate the species tree from gene trees. Summary methods operate by combining estimated gene trees and thus suffer in the presence of low phylogenetic signal. To tackle this problem the Statistical Binning and Weighted Statistical Binning pipelines were designed to improve gene tree estimation, which in turn can improve species tree estimation. Experimental studies of these pipelines revealed that they helped in improving gene tree and species tree estimation. However, these studies only tested the weighted statistical binning and statistical binning pipelines using multi-locus bootstrapping (MLBS) and not using BestML, where MLBS and BestML are different ways to run a phylogenetic pipeline. 
  In this thesis, a novel phylogenetic pipeline named WSB+WQMC is proposed. This pipeline shares several design features with the weighted statistical binning pipeline (referred as WSB+CAML in this thesis) but has some other desirable properties. The WSB+WQMC pipeline is also shown to be statistically consistent under the GTR+MSC model when a slightly different version of WQMC is used.  
  
  In this study WSB+WQMC was evaluated and compared with the WSB+CAML pipeline on various simulated datasets using BestML analysis. Most of the trends seen in MLBS analyses were also observed for WSB+WQMC and WSB+CAML in BestML analyses with some important differences. It is shown that WSB+WQMC substantially improved the accuracy of gene tree and species tree estimation using ASTRAL2 and ASTRID on most datasets having low, medium, and moderately high levels of ILS. Compared to WSB+CAML, it was found that WSB+WQMC computed less accurate gene trees and species trees in certain model conditions having low and medium levels of ILS. However, WSB+WQMC was found to be better and at least as accurate as WSB+CAML in computing gene trees and species trees on all datasets having moderately high and high ILS levels. WSB+WQMC is also shown to be better in estimating gene trees on certain medium and low ILS datasets. Thus, WSB+WQMC is a potential alternative to WSB+CAML for gene tree and species tree estimation in the presence of low phylogenetic signal.
\end{doublespace}
\newpage
\section*{ACKNOWLEDGMENTS}
\begin{doublespace}
I would like to express profound gratitude to my advisor Dr. Tandy Warnow for giving me
the opportunity to be a part of this project and all the guidance and support she has given me
along the way. I am inspired by the high standards that she sets for herself and those around her, her hard work, her attention to detail, her dedication to the profession, and her intense commitment to her work. I
have thoroughly enjoyed working with her and feel like I have learned a lot. 

I would like to thank my fellow lab mates: Nam Nguyen, Mike Nute, Pranjal Vachaspati, Erin Molloy, Jed Chou, and Ruth Davidson, for the stimulating discussions and for all the fun we have had in the last one year. I would also like to thank Dr. Siavash Mirarab and Md. Shamsuzzoha Bayzid for their help in my experiments. 

Finally, I must express my very profound gratitude to my parents for providing me with unfailing support and continuous encouragement throughout my years of study and through the process of researching and writing this thesis. This accomplishment would not have been possible without them.

\end{doublespace}
\newpage
\tableofcontents
\newpage
\mainmatter
\chapter{Introduction}
\begin{doublespace}
\section{Motivation}
    %Problems
        %concatenation used for getting supergene trees 
            %large running time
            %true gene tree still have discord
        % initially estimated gene tree is not explicitly used to compute new gene tree
        %Oly tested in MLBS analysis

Phylogenetic trees, whether of individual loci (so called ``gene trees") or genome-level (species trees) provide a basis for understanding how life evolved on earth and have many biological and medical applications.  Thus, the estimation of gene trees and species trees is a basic step in many biological analyses \cite{eisen1998phylogenomics}. However, computing these trees with high accuracy is very challenging due to a variety of reasons including computational challenges (almost all problems are NP-Hard) and limited amounts of data. Due to limited amount of data for individual loci, the gene alignments used in phylogenetic studies often have low phylogenetic signal (i.e., its sequences are too short or it evolves too slowly) \cite{bayzid2013naive,mirarab2014statistical,bayzid2015weighted}. Gene tree estimation together with low phylogenetic signal is also plagued by errors in gene alignment itself. Moreover, many biological conditions (e.g., short branches, long branches, etc.) also make it impossible to estimate highly accurate gene trees.

Species tree estimation on the other hand faces the problem of individual loci having different evolutionary histories, caused by the biological phenomenon known as incomplete lineage sorting (ILS) \cite{maddison1997gene,degnan2009gene,degnan2006discordance,edwards2009new}. Furthermore, in the presence of ILS, standard methods like concatenation and consensus can be statistically inconsistent \cite{roch2015likelihood,degnan2009properties} and are known to perform poorly under some conditions \cite{kubatko2007inconsistency}. Therefore, many new methods have been developed for estimation of species trees in the presence of ILS (e.g., co-estimation methods, summary methods, etc.). Co-estimation methods, which co-estimate gene trees and species tree like BEST \cite{liu2008best} and *BEAST \cite{drummond2007beast}, generally have excellent accuracy but are limited by their computational complexity. Summary methods on the other hand operate by combining estimated gene trees into a species tree. Some of the summary methods like ASTRAL2 \cite{mirarab2015astral}, ASTRID \cite{vachaspati2015astrid} and MP-EST \cite{liu2010maximum} are based on the multi-species coalescent model \cite{kingman1982coalescent} and are statistically consistent under that model. Summary methods that are statistically consistent under the multi-species coalescent model reconstruct the true species tree with high probability given a sufficiently large number of true gene trees. Summary methods are the most frequently used methods for estimating species trees under the multi-species coalescent model, and have shown good results on many biological datasets \cite{liu2007species,liu2008best}.

Summary methods assume perfectly accurate gene trees for statistical consistency and thus suffer when the gene trees have low accuracy \cite{mirarab2014astral, leache2010accuracy}. Thus, an attempt to improve gene trees by enhancing phylogenetic signal not only helps in getting better gene evolutionary histories, but it also indirectly affects the accuracy of species tree computed by summary methods. One obvious solution to the problem is collecting more data, which is both costly and time consuming, making it infeasible for most studies. If collecting more data is not an option, another obvious direction to boost the phylogenetic signal of an individual gene is by using the data from other genes. Co-estimation methods like BEST and *BEAST use this idea to enhance gene tree and species tree estimation. However, their huge computational complexity limits their use in most studies.

The Naive Binning pipeline \cite{bayzid2013naive} was another direction for boosting phylogenetic signal and showed improvements in the accuracy of estimated species trees. In this pipeline genes are randomly grouped into bins of same size and a supergene alignment is computed for each bin by concatenating gene alignments present in that bin. The naive binning pipeline then uses maximum likelihood to get a supergene tree on each bin. These supergene trees can then be used to compute a species tree using a summary method. Statistical Binning \cite{mirarab2014statistical} and Weighted Statistical Binning \cite{bayzid2015weighted} improved the naive binning pipeline and grouped genes in a smarter way to ensure genes within a bin have similar evolutionary histories and less discord. Both the statistical binning and the weighted statistical binning pipelines were evaluated on various datasets and showed positive results. Weighted statistical binning was also proved to be statistically consistent under the GTR+MSC model, a model where gene trees evolve within a species tree under the multi-species coalescent (MSC) model \cite{kingman1982coalescent} and sequences evolve down each gene tree under the General Time Reversible (GTR) model \cite{tavare1986some}. However, these pipelines were only tested using multi-locus bootstrapping (MLBS) analysis of datasets and not using BestML analysis, which is known to be more accurate than MLBS in certain model conditions having large enough number of genes \cite{mirarab2014evaluating}. Second, statistical binning and weighted statistical binning used concatenation within each bin, which is not only computationally expensive but may also lead to poor gene tree accuracy if gene trees within a bin have different evolutionary histories. Third, statistical binning and weighted statistical binning discarded the topology of the initially estimated gene tree, which potentially can be fairly accurate and close to the true evolutionary history in many cases. Finally,  statistical binning and weighted statistical binning computed a single supergene tree for each bin rather than a unique gene tree for each input gene tree. 

In this thesis a novel phylogenetic pipeline named WSB+WQMC is proposed. It aims to improve gene tree and species tree estimation when the individual loci have low phylogenetic signal. It modifies the weighted statistical binning pipeline and computes a new unique gene tree for each initial gene tree. The new pipeline uses WQMC \cite{avni2015weighted} (a quartet-based tree estimation method) to compute a new gene tree from a set of weighted quartets unique to each gene. In the WSB+WQMC pipeline, quartets from initially estimated gene trees (i.e.,  original gene trees) within a bin are combined uniquely for each gene by up-weighting its own quartets; hence, it allows WSB+WQMC to use the topology of initially estimated gene tree to its advantage. The WSB+WQMC pipeline has good theoretical properties and is shown to be statistically consistent under the GTR+MSC model when a slightly different version of WQMC is used in the pipeline. The WSB+WQMC pipeline is tested on BestML analysis of various datasets and the results are compared with the original weighted statistical binning pipeline (which we refer to as WSB+CAML) proposed in \cite{bayzid2015weighted}. 
\section{Background}
\subsection*{Basic phylogenetic pipeline using a summary method (Unbinned)}
\begin{figure}[ht]
\centering
\includegraphics[width=\textwidth]{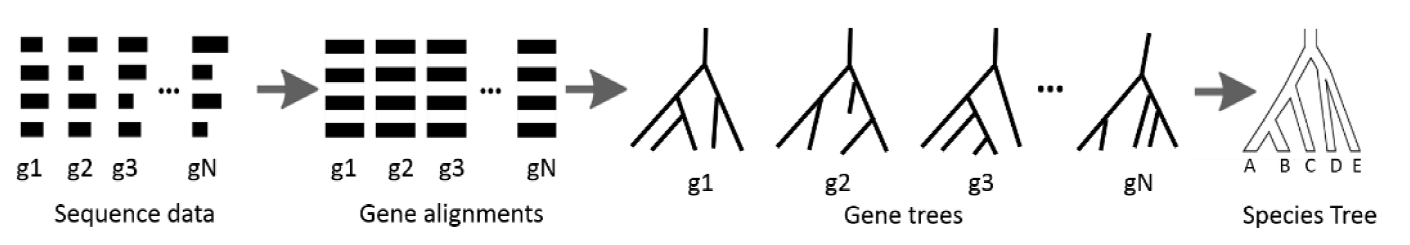}
\caption{\textbf{Basic unbinned phylogenetic pipeline using a summary method.} The input to the pipeline is a set of sequences for different loci across different species. In this pipeline, a multiple sequence alignment is computed using any alignment method. Then a gene tree is computed on each gene using the multiple sequence alignment. Gene trees are then used by a summary method to compute  a species tree. }
\label{fig:traditional-pipeline}
\end{figure}

The basic unbinned phylogenetic pipeline using a summary method is shown in Figure \ref{fig:traditional-pipeline}. The pipeline begins with a set of sequences
for different loci across different species as input. Gene alignments for each gene are then computed using any alignment method. Next, gene trees are estimated on each locus using the input gene alignments. Finally, it combines the estimated gene trees into a species tree using a summary method.  
\subsection*{MLBS vs. BestML}
Multi-locus bootstrapping (MLBS) and BestML are different ways to run a phylogenetic pipeline that uses a summary method. When running the pipeline using BestML analysis, a single gene tree having the best maximum likelihood score is estimated for each gene. Then, the best maximum likelihood gene tree estimate for each gene is used by a summary method to compute a species tree. 

Given $n$ genes having $k$ bootstrap replicated gene alignments each, the phylogenetic pipeline using MLBS analysis is run using the following steps.  
\begin{itemize}
\item For each gene $i$ and bootstrap replicated alignment $j$, a tree $t_{i,j}$ is estimated using a maximum likelihood method with $j^{th}$ bootstrap replicated gene alignment for gene $i$ as input. Here $1 \leq i \leq n$ and $1 \leq j \leq k$.
\item Then, for $j=1$ up to $k$, a summary method is run with the $j^{th}$ bootstrap replicated alignment for each gene as input (i.e., $\{t_{1,j},t_{2,j},...,t_{n,j}\}$ is the input to the summary method in $j^{th}$ run) to compute a species tree $t^{sp}_j$.   
\item Finally, a greedy consensus tree of $k$ species trees from the previous step (i.e., $\{t^{sp}_{1},t^{sp}_{2},...,t^{sp}_{k}\}$) is computed. This greedy consensus tree is the output species tree. 
\end{itemize}

Multi-locus bootstrapping (MLBS) is more frequently used in phylogenetic analyses. However, MLBS is not necessarily the best way to compute gene trees and species trees. In \cite{mirarab2014evaluating}, the accuracy of species tree estimation methods like Greedy consensus, MRP \cite{ragan1992phylogenetic}, MRL \cite{nguyen2012mrl}, and MP-EST \cite{liu2010maximum} using MLBS analysis was compared to the accuracy of species tree estimation of these methods using BestML analysis. It was found that all methods computed more accurate species tree using MLBS analysis when the number of genes used was small. On the other hand, when a large enough number of genes were used, it was found that all methods computed more accurate species tree using BestML analysis. This highlights the importance of testing and evaluating all phylogenetic pipelines using both MLBS and BestML analyses.  

\subsection*{Naive Binning}
The benefit of using binning-based pipelines to improve gene tree estimation has only recently been discovered. The first study that explored binning was \cite{bayzid2013naive}, which introduced the naive binning pipeline. In this technique, genes were randomly partitioned into disjoint bins and supergene alignments for each bin were computed by concatenating individual gene alignments present in that bin. Maximum likelihood trees were computed for each supergene alignment, and then these supergene trees were combined into a species tree using a summary method. 

The naive binning pipeline was evaluated using BestML analysis on simulated datasets with species tree estimation methods including *BEAST, concatenation, and several summary methods. It was observed that in low ILS conditions, the naive binning pipeline improved the accuracy of species tree estimation using summary methods and improved scalability of *BEAST without impacting its accuracy. These observations showed that naive binning can help in improving species tree estimation when the individual genes have low phylogenetic signal. However, since naive binning puts genes into bins randomly and can lead to genes with very different evolutionary histories put in the same bin, it was conjectured that it could reduce accuracy in high levels of ILS. 
\subsection*{Statistical Binning}
Statistical Binning \cite{mirarab2014statistical} improved on the naive binning pipeline and used a different binning technique than random binning. The new binning technique used bootstrap support values on the estimated gene trees to partition the set of loci into bins. The binning technique created bins of approximately equal sizes and ensured that no two genes within the same bin have conflicting edges having high support values in their respective tree topologies. This smarter binning technique ensured that genes with different evolutionary histories were unlikely to be placed in the same bin. Similar to naive binning, supergene alignments were computed for each bin by concatenating gene alignments within that bin. Then, supergene trees computed on the supergene alignments using a maximum likelihood gene tree estimation method were used to compute a species tree using a summary method. 

The statistical binning pipeline was evaluated in \cite{mirarab2014statistical} on many biological and simulated datasets using MLBS analysis and MP-EST as the summary method. It was observed that this pipeline helped in reducing error in the MP-EST species tree topology on all the datasets studied. It was also observed that the relative improvement in the accuracy of MP-EST species tree with respect to the basic unbinned pipeline was more when the level of ILS was low. However, the statistical binning pipeline was only tested on datasets having 37 or more taxa and using only MP-EST as the summary method. Moreover, statistical binning wasn't evaluated on datasets having ILS levels more than 60\% AD. In \cite{mirarab2014statistical}, the theoretical properties of statistical binning were also not addressed. Subsequently, Theorem 3 from \cite{bayzid2015weighted} showed that the statistical binning pipeline is statistically inconsistent under the GTR+MSC model.
\subsection*{Weighted Statistical Binning (WSB+CAML)}
\label{section:wsbcaml}
\begin{figure}[p]
\centering
\includegraphics[width=\textwidth]{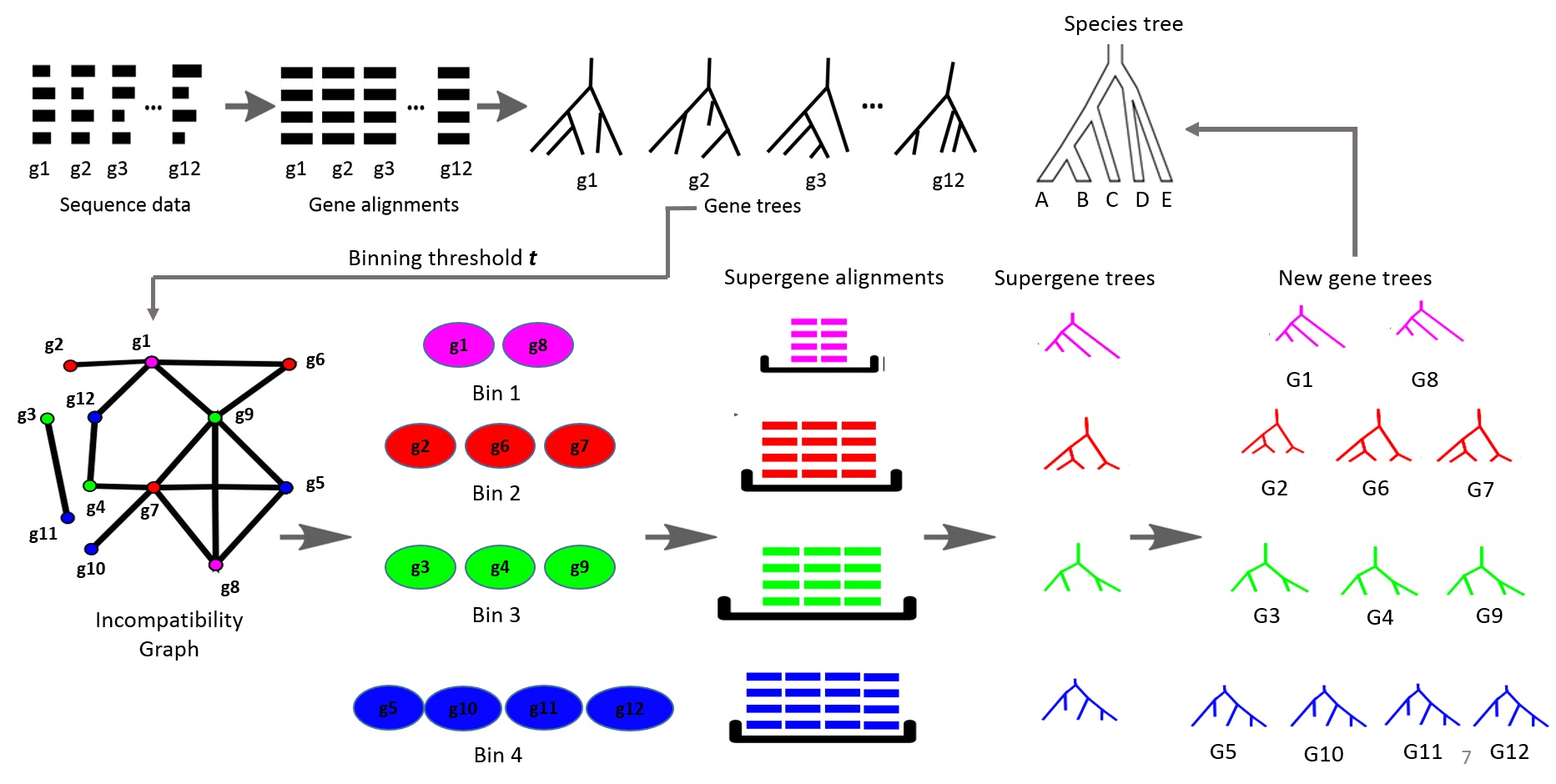}
\caption{\textbf{WSB+CAML pipeline for phylogenetic analysis.} The input to the pipeline is a set of gene alignments computed from a set of sequences for different loci across different species. Gene trees are then estimated on the gene alignments using a maximum likelihood tree estimation method. Gene trees are then used to compute an incompatibility graph, where each vertex represents a gene and each edge represents incompatibility between them based on binning threshold $t$. A heuristic for balanced minimum vertex coloring is run to divide the genes into disjoint bins. For each bin, gene alignments for genes within that bin are concatenated into a supergene alignment. Supergene trees are then estimated using a fully partitioned maximum likelihood gene tree estimation method. Supergene tree for each bin is repeated for as many genes in that bin and considered as the new gene tree. New gene trees are used by a summary method to compute a species tree. Please note that the example used to describe the WSB+CAML pipeline is taken from Figure 1 of \cite{bayzid2015weighted}.}
\label{fig:wsbcaml}
\end{figure}

The weighted statistical binning pipeline \cite{bayzid2015weighted} (referred as WSB+CAML in this thesis) modified the statistical binning pipeline by adding one more step before running a summary method on supergene trees. It repeated the supergene trees as many times as the number of genes in its bin, and used those repeated supergene trees as the input for a summary method to compute a species tree. Detailed steps in the weighted statistical binning pipeline are shown in Figure \ref{fig:wsbcaml}. The input to the WSB+CAML pipeline is a set of gene alignments on different loci. The WSB+CAML pipeline then proceeds with the following steps.
\begin{itemize}
    \item Step 1: It computes a maximum likelihood (ML) gene tree with bootstrap support values for each gene from its sequence alignment. These gene trees are also referred as initial or original gene trees.
    \item Step 2 (Statistical Binning): It computes an incompatibility graph based on a bootstrap threshold $t$, with each gene corresponding to a vertex in the graph. For each pair of vertices in the graph, an edge is added between them if the gene trees corresponding to the vertices are incompatible with each other. To determine incompatibility between gene trees, first all the edges having bootstrap support values less than $t$ are collapsed, and then the compatibility is checked between the edge-collapsed gene trees. Two gene trees are said to be compatible with each other if they have a common refinement. After that, a heuristic from \cite{mirarab2014statistical} is used to color the vertices of the incompatibility graph such that no two adjacent nodes have the same color, the number of distinct colors used is small, and the color classes are approximately of the same size (minimum balanced vertex coloring problem). The coloring of the nodes defines the output bins with genes having same color in the same bin. 
    \item Step 3: For each bin, gene alignments of its constituent genes are concatenated and define the supergene alignment. 
    \item Step 4: A supergene tree is then estimated for each supergene alignment using a fully partitioned maximum likelihood gene tree estimation method.
    \item Step 5: Each supergene tree is repeated as many times as the number of genes in its bin. For each gene, the supergene tree for its bin is considered as its new gene tree. 
    \item Step 6: The new gene trees computed in step 5 are used as input to estimate a species tree using a summary method.
\end{itemize}
The WSB+CAML pipeline was found to have strong theoretical guarantees. Theorems 1 and 2 from \cite{bayzid2015weighted} showed that this small change in the statistical binning pipeline of repeating supergene trees made the new weighted statistical binning pipeline statistically consistent under the GTR+MSC model.\\\\
\textbf{Theorem 1 from \cite{bayzid2015weighted}}: Let $T^{sp}$ be a species tree with branch lengths in coalescent units, and $\mathcal{T} = \{t_1,t_2,...,t_p\}$ be a set of $p$ rooted gene trees sampled from the distribution defined by $T^{sp}$ under the multi-species coalescent model. Let $\{\theta_1, \theta_2,...,\theta_p\}$ be a set of numeric GTR model parameters (gene tree branch lengths and 4 $\times 4$ substitution matrices) so that $T_i = (t_i,\theta_i)$ is a GTR model tree for each $i=1,2,...,p$. Let $\mathcal{T'} = \{T_1,T_2,...,T_p\}$. For each $i$, $1 \leq i \leq p$, let sequence dataset $S_i$ evolve down the GTR model tree $T_i$. Let $\epsilon < 1$ and bootstrap support threshold $B < 1$ be given. Then, there is a sequence length L (that depends on $\mathcal{T'}$ and $\epsilon)$ such that if at least $L$ sites evolve down each gene tree, then with probability at least $1-\epsilon$, the following will be true. 
\begin{itemize}
    \item For each $i=1,2,..p$, the gene tree estimated using a GTR maximum likelihood method on $S_i$ will have the same unrooted topology as $t_i$ (the true gene tree for $S_i$) and will have bootstrap support values greater than $B$ for all its branches. 
    \item For every bin produced by the binning technique used in WSB+CAML based on GTR maximum likelihood analyses of the gene sequence alignments, the estimated gene trees for genes in the bin will have the same topology, and
    \item All genes with the same true gene tree topology will be in the same bin.  
\end{itemize}
This theorem is copied directly from \cite{bayzid2015weighted} and presented here as it is later used in the statistical consistency proof of WSB+WQMC.\\\\
\textbf{Theorem 2 from \cite{bayzid2015weighted}}: The phylogenetic pipeline that uses GTR maximum likelihood to estimate gene trees, uses weighted statistical binning to compute supergene trees, and then combines the supergene trees using a coalescent-based summary method, is statistically consistent under the GTR+MSC model.

Note that the term ``coalescent-based summary method" in this theorem refers to summary methods that are statistically consistent under the multi-species coalescent model.\\

In \cite{bayzid2015weighted}, the weighted statistical binning pipeline was evaluated and compared with the statistical binning pipeline on various biological and simulated datasets using MLBS analysis and two summary methods, ASTRAL2 and MP-EST. It was found that statistical binning and weighted statistical binning had no significant differences in terms of species tree estimation on the datasets studied. It was also found that weighted statistical binning helps in improving species tree topologies except for a small number of datasets having very high levels of ILS and small numbers of taxa. The relative magnitude of improvement in the accuracy of species trees compared to the basic unbinned pipeline was also seen to be more for conditions having low ILS. The difference between species tree estimation error by running the WSB+CAML pipeline and the basic unbinned pipeline decreased with increasing
sequence length.  
\section{WSB+WQMC}\label{section:wsbwqmc}
\begin{figure}[p]
\centering
\includegraphics[width=\textwidth]{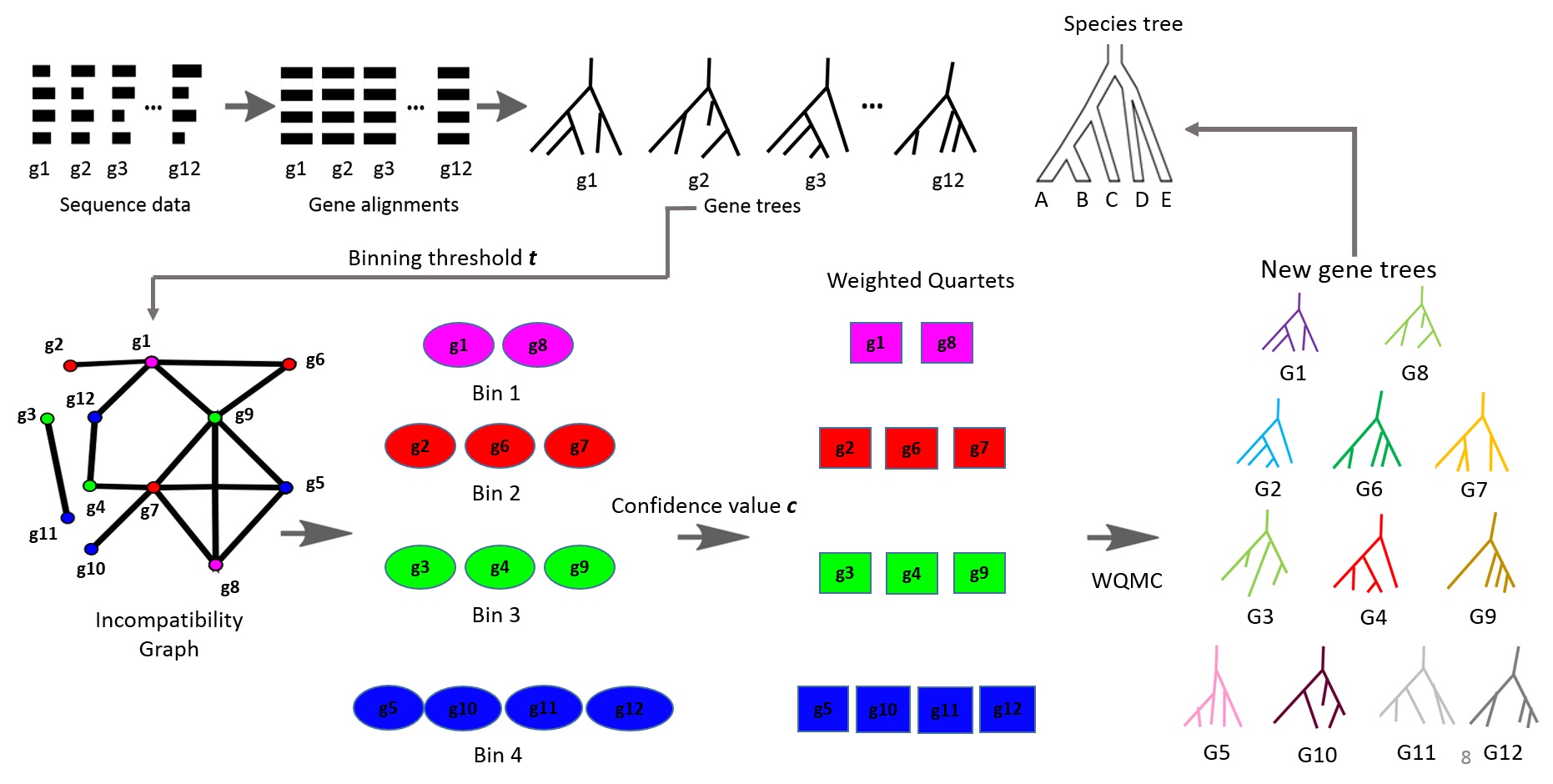}
\caption{\textbf{WSB+WQMC pipeline for phylogenetic analysis.} The input to the pipeline is a set of gene alignments computed from a set of sequences for different loci across different species. Gene trees are then estimated on the gene alignments using a maximum likelihood tree estimation method. Then, genes are divided into disjoint bins using the binning technique used in WSB+CAML with binning threshold $t$. For each gene, a set of weighted quartets is computed by combining weighted quartet topologies within its bin with up-weighting own quartets by confidence value $c$. WQMC is then run with weighted quartet set as input to get a new gene tree for each gene. New gene trees are then used by a summary method to compute the species tree.}
\label{fig:wsbwqmc}
\end{figure}

In this section, a novel phylogenetic pipeline named WSB+WQMC is presented, which aims to improve the accuracy of gene tree and species tree estimation compared to the weighted statistical binning pipeline. This pipeline modifies the WSB+CAML pipeline and avoids computing supergene alignments and supergene trees. Rather, it computes a set of weighted quartets for each gene using quartets induced by its original gene tree and other quartet topologies from genes within its own bin. It computes a unique weighted quartet set for a gene by combining the weighted quartets induced by other original gene trees within its bin and weighted quartets induced by itself (up-weighted by a factor of confidence value $c$ $\times$ size of bin). These weighted quartet sets are then used to compute the new gene trees using WQMC \cite{avni2015weighted}. These new gene trees are then used to estimate a species tree by a summary method. Detailed steps in the WSB+WQMC pipeline are shown in Figure \ref{fig:wsbwqmc}. The input to the WSB+WQMC pipeline is a set of gene alignments on different loci. The WSB+WQMC pipeline then proceeds with the following steps.
\begin{itemize}
    \item Step 1: Same as step 1 in the WSB+CAML pipeline.

    \item Step 2 (Statistical Binning): Same as step 2 in the WSB+CAML pipeline. 
    \item Step 3: Given a confidence value $c$ (algorithmic parameter), for each gene, a set of weighted quartets is computed. Let $\mathcal{G}$ be the set of original gene trees and $B_1, B_2, .. B_k$ be the bins computed in step 2, where k is the number of bins. \\\\
    Consider a gene $g_1 \in \mathcal{G}$, WLOG $\{g_1,g_2,...,g_{|B_l|}\} = B_l$.
    \begin{itemize}
        \item[--] First the set $\mathcal{Q}_{g_1} = \{ \langle q,upweight*w_{g_1}(q) \rangle | \;q$ is a quartet topology induced by gene tree $g_1\}$ is computed. Here $w_{g}(q) = \frac{length \; of \; internal \; branch \; of \; quartet \; q \; induced\; by \;g}{diameter \; of \; the \; quartet \; q}$, and $upweight = c *|B_l|$, where confidence value $c$ is an algorithmic parameter. Note that the maximum likelihood tree comes with branch lengths that are non-negative and hence leaf-to-leaf distances are defined by the branch lengths. The diameter of a quartet is the largest distance between any two leaves in the tree. The set $\mathcal{Q}_{g_1}$ has a tuple (quartet topology and its weight) for all  the quartet topologies the gene tree $g_1$ induces.   
        \item[--] Second, for each other gene tree in the bin, $t \in \{B_l \setminus g_1 \}$, the set $\mathcal{Q}_t = \{\langle q,w_t(q) \rangle | \;q $ is a quartet topology induced by the gene tree t$\}$ is computed. 
        \item[--] Finally, all the sets $\mathcal{Q}_{g_1},\mathcal{Q}_{g_2},...\mathcal{Q}_{g_{|B_l|}}$ are merged into a single weighted quartet set $\mathcal{M}_{g_1}$, where the weight of any quartet topology is the sum of weights of the same quartet topology in all other sets. 
    \end{itemize}
These steps are repeated to obtain all weighted quartet sets $\mathcal{M}_{g_1},\mathcal{M}_{g_2},...\mathcal{M}_{{|\mathcal{G}|}}$.\\\\
The weight function $w$ used to determine the weight of the quartet was used in \cite{avni2015weighted} and is a measure of reliability of the quartet. 
\item Step 4: The new gene tree is computed for each gene by running WQMC on its weighted quartet set as input. 
\item Step 5:  The new gene trees computed in step 4 are used as input to estimate a species tree using a summary method.
\end{itemize}

The basic intuition behind WSB+WQMC is to use the frequent quartet topologies present in the bin to possibly rectify incorrect quartet topologies induced by an original gene tree. The confidence value $c$ used in the pipeline determines the robustness to change from the original gene tree topology. The larger the confidence value c, the more frequent a conflicting quartet topology must occur in the bin to change the gene tree topology. Thus, by having this control over robustness to change, a user can control the extent of deviation of new gene tree from the original gene tree. The binning threshold value $t$ used in step 2 on the other hand affects the size of bins. A large binning threshold results in larger bin sizes with only a single bin for $t=100\%$ (referred as one-bin), whereas a smaller binning threshold results in smaller bin sizes with singleton bins for $t=0\%$. Determining both the optimal binning threshold $t$ and confidence value $c$ is a very challenging problem and affects the performance of the pipeline.
\subsection*{Proof of statistical consistency of WSB+WQMC when WQMC used in the pipeline is replaced by WQMC*}
In this section, it is proved that WSB+WQMC is statistically consistent under the GTR+MSC model when a slightly different version of WQMC is used in the pipeline. For the purpose of this proof, WQMC used in the WSB+WQMC pipeline is replaced by WQMC*. WQMC* is a slightly different version of WQMC with an additional property that it exactly solves the trivial maximum weighted quartet compatibility (MWQC) problem (defined below). The statistical consistency proof given in this section is in the same lines as the statistical consistency proof described in \cite{bayzid2015weighted} for WSB+CAML. The WSB+WQMC pipeline uses the same binning technique as used in WSB+CAML and therefore Theorem 1 from \cite{bayzid2015weighted} also applies to WSB+WQMC. The main result of this section is described in Theorem 2. The trivial MWQC problem is solvable in polynomial time using the All Quartets Method (defined below) and therefore replacing WQMC with WQMC* in the WSB+WQMC pipeline for the purpose of this proof is reasonable.\\\\
\textbf{Definitions}\\
 \textbf{Maximum weighted quartet compatibility problem (MWQC)}: Given a set $Q = \{\langle q, w(q) \rangle | \; q$ is a quartet topology and $w$ is any positive weight function$\}$, find a tree $t$ that maximizes the weight $W(t)$, where $W(t) = \sum\limits_{ x \in Q'(t)} w(x) $, where $Q(t)$ is a set containing all the quartet topologies induced by the tree $t$, and $Q'(t) = Q \cap Q(t)$.\\\\
 \textbf{Trivial weighted quartet set:} A weighted quartet set $Q= \{\langle q, w(q) \rangle | \; q$ is a quartet topology and $w$ is any positive weight function$\}$ is called a trivial weighted quartet set when it contains all quartet topologies induced by a tree t and nothing else.\\\\
 \textbf{Trivial MWQC problem}: An instance of MWQC problem is said to be trivial when the set $Q$  is a trivial weighted quartet set from a tree $t^*$. It can be easily seen that the solution to the trivial MWQC problem is the tree $t^*$ itself\\\\
 \begin{mdframed}
 \textbf{All Quartets Method} (This is taken from \cite{warnow2016book})\\The input to the method is a weighted quartet set $Q$. Let $S=\{s_1,s_2,...,s_n\}$ be the leaf set of the quartet trees in $Q$. It is assumed that $|S| \geq 4$, since otherwise there are no quartets. Also, it is assumed that the quartet set $Q$ contains exactly one quartet tree on every four leaves. Otherwise, the All Quartet Method fails to compute a tree and returns ``No compatibility tree". Given the input set $Q$, the All Quartets Method proceeds with the following steps.
 \begin{itemize}
     \item if $|S| = 4$, then return the quartet tree in $Q$. Else, find a pair $s_i,s_j$ that are always grouped together in any quartet that includes both $s_i,s_j$. If no such pair exists, return ``No compatibility tree" and exit. Otherwise, remove all the quartets that include both $s_i,s_j$ from the set $Q$.  
     \item Recursively compute a tree $t'$ on $S - \{s_i\}$.
     \item Return the tree created by inserting $s_i$ next to $s_j$ in $t'$. 
 \end{itemize}
 \end{mdframed}
 \textbf{Note}: It can be easily seen that the All Quartets Method returns the tree $t$, when the trivial weighted quartet set from a tree $t$ is given as input.\\\\
 \textbf{WQMC*}: WQMC* is a slightly different version of WQMC and proceeds as follows. 
 \begin{itemize}
     \item Given an input set of weighted quartets $Q$, the All Quartets Method is run with the set $Q$ as input. If the All Quartets Method returns a compatibility tree $t$, return $t$ and exit. Otherwise, proceed to the next step.  
     \item Run WQMC with the set $Q$ as input and return the output tree. 
 \end{itemize}
\textbf{Corollary 1:} Let $T^{sp}$ be a species tree with branch lengths in coalescent units, and $\mathcal{T} = \{t_1,t_2,...,t_p\}$ be a set of $p$ rooted gene trees sampled from the distribution defined by $T^{sp}$ under the multi-species coalescent model. Let $\{\theta_1, \theta_2,...,\theta_p\}$ be a set of numeric GTR model parameters (gene tree branch lengths and 4 $\times 4$ substitution matrices) so that $T_i = (t_i,\theta_i)$ is a GTR model tree for each $i=1,2,...,p$. Let $\mathcal{T'} = \{T_1,T_2,...,T_p\}$. For each $i$, $1 \leq i \leq p$, let sequence dataset $S_i$ evolve down the GTR model tree $T_i$. Let $\epsilon < 1$ and bootstrap support threshold $B < 1$ be given. Then, there is a sequence length L (that depends on $\mathcal{T'}$ and $\epsilon)$ such that if at least $L$ sites evolve down each gene tree, then with probability at least $1-\epsilon$, the following will be true. 
\begin{itemize}
    \item For each $i=1,2,..p$, the gene tree estimated using a GTR maximum likelihood method on $S_i$ will have the same unrooted topology as $t_i$ (the true gene tree for $S_i$) and will have bootstrap support values greater than $B$ for all its branches. 
    \item For every bin produced by the binning technique used in WSB+WQMC based on GTR maximum likelihood analyses of the gene sequence alignments, the estimated gene trees for genes in the bin will have the same topology, and
    \item All genes with the same true gene tree topology will be in the same bin.  
\end{itemize}
\textbf{Proof:} The bins produced by WSB+CAML and WSB+WQMC are identical. Therefore, we have this corollary that is identical to Theorem 1 from \cite{bayzid2015weighted} except that WSB+CAML is replaced by WSB+WQMC. \\\\
\textbf{Theorem 2:} The WSB+WQMC phylogenetic pipeline with WQMC replaced by WQMC* is statistically consistent under the GTR+MSC model.\\
\textbf{Proof:} By Corollary 1, as the sequence length for each gene tree goes to infinity, all the genes put in any bin by WSB+WQMC will have the same true gene tree topology with probability converging to 1. This will cause the weighted quartet sets for each gene $g_i$ to be a trivial weighted quartet set with quartet topologies only from the true gene tree $t_i$. Since it has been assumed that WQMC* used to compute a gene tree from a weighted quartet set solves the trivial MWQC exactly, the new gene trees will have the same topology as their corresponding true gene tree. Hence, the distribution of new gene trees produced using WSB+WQMC will be identical to the distribution of the true gene trees for these genes. Therefore, as number of sites and number of genes increase, the gene tree distribution from WSB+WQMC gene trees converges to the true gene tree distribution. Finally, as the gene tree distribution converges to the true gene tree distribution, the species tree computed by a statistically consistent summary method also converges to a true species tree.  

\newpage
\section{Figures and Tables}
\vfill
\vfill
\clearpage
\newpage
\end{doublespace}
\chapter{Experimental Study}
\begin{doublespace}
\section{Datasets}
In the experimental study, datasets were separated into two disjoint sets: a training dataset and a testing dataset. Table \ref{table:training} lists the datasets used for training and Table \ref{table:testing} lists the datasets used for testing. The ILS level in the datasets was measured using AD\%, which is the average percentage of missing branches between true gene trees and the true species tree. Higher value of AD\% indicates higher level of ILS in the dataset. For the purpose of this study, the range of ILS levels was broken down into four different categories, namely low, medium, moderately high, and high, as shown in Table \ref{table:ils}. Note that the high level of ILS has a very big range and there are probably huge differences between the low end of the high ILS range and the high end of the high ILS range. 

The mammalian simulated dataset studied in \cite{mirarab2014statistical} (based on an MP-EST \cite{liu2010maximum} analysis of a mammalian biological dataset having 37 taxa) was used for the training phase. For the testing phase, collections of simulated datasets with 10, 11, 15, and 50 taxa along with the Mammalian-1X dataset having 250bp alignment length were used. The ILS level in the simulated datasets used in the study ranged from low ILS (11-taxon L ILS and SimPhy-10M10e7) to high ILS conditions (10-taxon H ILS, 15-taxon H ILS, SimPhy-500K10e7, and SimPhy-500K10e6). All the datasets except the SimPhy datasets had bootstrap support values on the original gene trees available from earlier studies. The 15-taxon datasets evolve under a strict molecular clock, but the other datasets do not. 

The original gene trees for the Mammalian, 10-taxon, 15-taxon and 11-taxon datasets were computed in prior studies \cite{bayzid2015weighted,mirarab2014statistical,bayzid2013naive} from the gene alignments using maximum likelihood estimation method RAxML version 7.3.5 \cite{stamatakis2006raxml}. To compute the original gene trees from gene alignments, the maximum likelihood method FastTree 2.1 \cite{price2010fasttree} was used for the SimPhy datasets. These maximum likelihood methods were also used to estimate supergene trees from the supergene alignments in the WSB+CAML pipeline. RAxML version 8.2.7 was used in unpartitioned mode for computing  supergene trees from the supergene alignments for the mammalian dataset. For the 10-taxon, 11-taxon and 15-taxon datasets, RAxML version 8.2.7 was used in fully partitioned mode to compute supergene trees from the supergene alignments. To compute supergene trees from the supergene alignments for the SimPhy dataset, FastTree 2.1 was used in unpartitioned mode. More detailed information about each dataset is presented in the following paragraphs.
\subsection{Mammalian}
This dataset was also studied in \cite{mirarab2014statistical}. This dataset is based on an MP-EST analysis of a mammalian biological dataset and has 37 taxa. To vary the amount of ILS, the branch lengths in the species tree estimated using MP-EST on biological data were modified. To simulate higher level of ILS the branch lengths were shortened by dividing by 2 (Mammalian-0.5X) and to simulate lower level of ILS branch lengths were multiplied by 2 (Mammalian-2X). For the Mammalian-1X datasets, the branch lengths were kept unchanged. This datasets ranges in ILS level from low (21\% AD) for the 2X branch length to high (50\% AD) for 0.5X branch length. The Mammalian-1X dataset provides multiple sequence alignment of lengths 250bp and 500bp. This dataset doesn't follow a strict molecular clock.

The dataset already contained gene trees estimated from the sequence alignments with bootstrap values using RAxML version 7.3.5. The estimated gene trees already available in these datasets were used as the original gene trees in this study.
\subsection{10-taxon}
 This dataset was previously studied in \cite{bayzid2015weighted} and generated using SimPhy \cite{mallo2015simphy}. This dataset provides two different ILS conditions, the first having moderately high ILS (40\% AD) and second having high ILS (84\% AD). This dataset is very heterogeneous with a different species tree for each replicate and various genes having different rates of evolution. The average bootstrap support for this dataset ranged from 45\% for the high ILS condition to 34\% for the moderately high ILS condition. This dataset doesn't follow a strict molecular clock.

The dataset already contained gene trees estimated from the sequence alignments with bootstrap values using RAxML version 7.3.5. The estimated gene trees already available in these datasets were used as the original gene trees in this study.
\subsection{15-taxon}
This dataset was simulated and studied in \cite{bayzid2015weighted}. This dataset has high ILS level (82\% AD). This dataset is very homogeneous with each replicate having the same caterpillar species tree. Gene trees from this dataset follow a strict molecular clock and differ from each other only due to the multi-species coalescent process. The ultrametric gene trees were simulated using McCoal \cite{yang2015mccoal} for this dataset. In the simulation process, sequences of length 1000bp were simulated down the gene trees and then 100bp were sampled to vary the number of sites. The average bootstrap support for the 100bp condition was 35\%, whereas the average bootstrap support for the 1000bp condition was 70\%. 

The dataset already contained gene trees estimated from the sequence alignments with bootstrap values using RAxML version 7.3.5. The estimated gene trees already available in these datasets were used as the original gene trees in this study.
\subsection{11-taxon}
This dataset was initially developed for \cite{chung2011comparing}, and later studied in \cite{bayzid2013naive}. Between the two 11-taxon datasets available from \cite{bayzid2013naive}, the dataset with more ILS was used in this thesis. The dataset used in this thesis had a low level of ILS (15\% AD). This dataset doesn't follow a strict molecular clock and is very heterogeneous with substantial rate variation among the gene trees and species tree. 100 gene multiple sequence alignments were available and used in this thesis with a length of 500bp. 

The dataset already contained gene trees estimated from the sequence alignments with bootstrap values using RAxML version 7.3.5. The estimated gene trees already available in these datasets were used as the original gene trees in this study.
\subsection{SimPhy}
This dataset was developed for \cite{mirarab2015astral} and simulated using SimPhy \cite{mallo2015simphy}. This dataset was simulated with 200 taxa, variable tree lengths (500K, 2M and 10M generation) and speciation rates (1e-6 and 1e-7 per generation). The tree length and speciation rate affects the amount of ILS, with shorter lengths resulting in shorter branches and therefore higher levels of ILS. Speciation rate impacts whether speciation events tend to happen close to leaves (1e-6) or close to the root (1e-7).

In this study, the sequence alignments for the first 50 taxa were sampled from the original multiple sequence alignment. Then, sequences of length 50bp, 100bp, and 200bp were sampled from the new sequence alignments having 50 taxa. Then, gene trees were estimated on these alignments using FastTree \cite{price2010fasttree} without bootstrap support values. FastTree is a maximum likelihood method for estimating gene trees from gene alignments. FastTree also produces local support values for each edge in the estimated gene tree. The different tree shapes used in the simulation (i.e., the tree length and the speciation rate) produced many model conditions with amount of ILS ranging from 9\% to 69\% AD. This dataset is very heterogeneous with a different species tree for each replicate and various genes having different rates of evolution. The sequences in this dataset do not evolve under a strict molecular clock. 
\section{Design}
In this experimental study, the WSB+WQMC pipeline was evaluated through a range of experiments on various simulated datasets that varied in many respects (sequence length of locus, number of taxa, number of genes, deviation from the molecular clock, and ILS level). The experiments were designed to answer the following questions.
\begin{itemize}
    \item The effect of WSB+WQMC on gene tree estimation error.
    \item The effect of WSB+WQMC on species tree estimation error when species trees are computed using either ASTRAL2 or ASTRID.
    \item The effect of algorithmic parameters in the WSB+WQMC pipeline.
    \item The relative performance of WSB+CAML and WSB+WQMC with respect to gene tree accuracy and species tree accuracy.
\end{itemize}

The experimental study in this paper was broken into two phases, namely the training and the testing phases. The datasets were also separated into two disjoint sets as shown in Tables \ref{table:training} and \ref{table:testing} for training and testing purposes.

\subsection{Training}
The training phase was required in this study to select the confidence value $c$ to be used in the testing phase.  In the training phase, the Mammalian-0.5X, Mammalian-1X and Mammalian-2X simulated datasets were analysed using different confidence values when running the WSB-WQMC pipeline. These datasets were chosen for the training phase because they provided a wide range of ILS levels (21\%, 30\%, and 50\% AD), and were used in \cite{bayzid2015weighted} to evaluate WSB+CAML. Based on the results on these training datasets, a single confidence value was chosen for the testing phase. Moreover, on these datasets, the effect of using two different binning thresholds (Binning 100\% and Binning 75\%) was also explored. 

The average bin sizes computed by the binning step in WSB+WQMC on the SimPhy-10M1e-7 dataset with different binning threshold values (Binning 90\% and Binning 95\%) were also analyzed in the training phase. This was done to select a single binning threshold value $t$ to be used for the SimPhy datasets in the testing phase. 

For all the testing datasets except the SimPhy datasets, no training was done to select a binning threshold value and binning threshold 75\% was used in the testing phase. This binning threshold value was used because the same binning threshold value was also used in \cite{bayzid2015weighted} for running WSB+CAML on these datasets. 

\subsubsection{Mammalian}
The WSB+WQMC pipeline on the Mammalian-0.5X, Mammalian-1X and Mammalian-2X datasets was run for different model conditions, as shown in Table \ref{table:training}. The input gene trees for the WSB+WQMC pipeline were BestML gene trees estimated using RAxML version 7.3.5 with 200 bootstrap replicates. For model conditions with 50 genes, the first 50 gene trees were sampled from 200 genes trees and used as input.

To explore the effect of confidence values, the WSB+WQMC pipeline was run multiple times once for each confidence value (0.0, 0.2, and 0.3) with binning threshold 75\%. Note that the confidence value of 0.0 leads to no up-weighting of genes and outputs the same new gene tree for all input gene trees within a bin. Another experiment to analyze the effect of binning threshold values was conducted by running the WSB+WQMC pipeline on the training datasets with binning threshold values 75\% and 100\% and confidence value 0.2. Binning 100\% puts all the gene trees into a single bin (also referred as one-bin in this study). The new gene trees produced by each run of the WSB+WQMC pipeline on each model condition were then used to compute a species tree using ASTRAL2 \cite{mirarab2015astral}. For analyzing the WSB+WQMC pipeline, the accuracy of the new gene trees computed by WSB+WQMC was compared to the accuracy of the original gene trees. Also, the accuracy of the ASTRAL2 species tree computed on new gene trees was compared to the accuracy of the ASTRAL2 species tree computed on original gene trees. More details on the evaluation of gene tree estimation and species tree estimation are provided in section \ref{section:methods}. 

\subsubsection{SimPhy}
The binning step in the WSB+WQMC pipeline was run on the SimPhy-10M1e-7 dataset having 50 taxa on model conditions having two different numbers of genes (100 and 200 genes) and three different alignment lengths (50bp, 100bp, and 200bp alignment length). For the model condition with 100 genes, the first 100 gene trees were sampled and used as input. For model conditions with sequence lengths 50bp and 100bp, the first 50 and 100 sites were sampled from the original alignment. 

The input gene trees were estimated from the sequence alignments using FastTree 2.1 without bootstrap support values. In order to run the binning step in  WSB+WQMC on this dataset, local support values computed by FastTree were used instead of bootstrap support values. It is important to note that the FastTree local support values are very different from bootstrap support values, and generally tend to be much higher than the bootstrap support values. To determine the binning threshold to be used in the testing phase, average bin sizes after the binning step in WSB+WQMC on the SimPhy-10M1e-7 dataset were computed with binning thresholds 90\% and 95\% (as shown in Table \ref{table:simphybins}). Only these very high binning threshold values were analyzed in the training phase as other lower binning threshold values would have resulted in singleton bins. It was found that binning threshold value 90\% produced substantially smaller bins than 95\%, making the latter a more preferable choice. Therefore, the binning threshold value of 95\% was chosen for running WSB+WQMC and WSB+CAML for the SimPhy datasets in the testing phase. 

\subsection{Testing}
Based on the results obtained for the training datasets (detailed in chapter \ref{chapter:results}), it was observed that the confidence values 0.2 and 0.3 were at par with each other and were a better choice than the confidence value 0.0. Specifically, the confidence values 0.2 and 0.3 were notably better than the confidence value 0.0 for gene tree estimation on all model conditions. Moreover, in terms of species tree estimation the confidence values 0.2 and 0.3 performed better in medium and high ILS conditions and were only worse in low ILS. Therefore, confidence value 0.2 was chosen as the only confidence value used in the rest of this study. For the Mammalian, 10-taxon, 15-taxon and 11-taxon datasets, the binning threshold value of 75\% was chosen to be used in the testing phase. For the SimPhy datasets, the binning threshold value of 95\% was chosen to be used in the testing phase. 

In this phase, experiments were conducted to analyze the performance of the WSB+WQMC pipeline on gene tree and species tree estimation using ASTRAL2 and ASTRID on the remaining datasets. Moreover, the WSB+CAML pipeline was also run on a subset of data and compared with the WSB+WQMC pipeline. For species tree estimation, ASTRAL2 and ASTRID were again used to compute species trees from gene trees.
\subsubsection{Mammalian}
In this phase, the WSB+WQMC pipeline was run on the Mammalian-1X dataset having 250bp alignment length, and the results obtained were compared with results from running WSB+WQMC on the Mammalian-1X dataset having 500bp alignment length. Moreover, the WSB+CAML pipeline was compared to the WSB+WQMC pipeline on the Mammalian-1X dataset having 200 genes and 250bp alignment length. Both the pipelines were run using binning threshold 75\%. The input gene trees for WSB+WQMC and WSB+CAML were BestML gene trees estimated from the gene alignments using RAxML version 7.3.5 with 200 bootstrap replicates. Supergene trees for each bin in the WSB-CAML pipeline were computed using BestML unpartitioned maximum likelihood analysis using RAxML version 8.2.7.
\subsubsection{10-taxon}
The WSB+WQMC pipeline was run on the 10-taxon H ILS and 10-taxon MH ILS datasets for various model conditions having different number of genes (shown in Table \ref{table:testing}). For model conditions with 25, 50, and 100 genes, the first 25, 50, and 100 gene trees were sampled from 200 genes trees respectively and used as input.

For each model condition, the WSB+WQMC pipeline was run using binning threshold 75\% and evaluated in terms of gene tree and species tree estimation. The WSB+CAML pipeline was also run on the 10-taxon MH ILS dataset having 100 genes and 100bp alignment length using binning threshold 75\% and compared with WSB+WQMC. The input gene trees for both WSB+WQMC and WSB+CAML were BestML gene trees estimated from gene alignments using RAxML version 7.3.5 with 200 bootstrap replicates. Supergene trees for each bin in the WSB-CAML pipeline were computed using BestML fully partitioned maximum likelihood analysis using RAxML version 8.2.7.
\subsubsection{15-taxon}
The WSB+WQMC pipeline was run on the 15-taxon H ILS dataset for different model conditions varying in number of genes and number of sites, as shown in Table \ref{table:testing}. For model conditions with 25, 50, 100, and 200 genes, the first 25, 50, 100, and 200 gene trees were sampled from 1000 genes trees and used as input. For the model condition with sequence length 100bp, the first 100 sites were sampled from the original alignment.  

For each model condition, the WSB+WQMC pipeline was run using binning threshold 75\%. The WSB+CAML pipeline was also run on the 15-taxon H ILS dataset having 100 genes and 100bp alignment length using binning threshold 75\% and compared with WSB+WQMC. The input gene trees for both WSB+WQMC and WSB+CAML were BestML gene trees estimated from gene alignments using RAxML version 7.3.5 with 200 bootstrap replicates. Supergene trees for each bin in the WSB-CAML pipeline were computed using BestML fully partitioned maximum likelihood analysis using RAxML version 8.2.7.
\subsubsection{11-taxon}
The WSB+WQMC pipeline was run on the 11-taxon dataset under different model conditions, as shown in Table \ref{table:testing}. The WSB+WQMC pipeline was run using binning threshold 75\%. For model conditions with 25 and 50 genes, the first 25 and 50 gene trees were sampled from 100 genes trees respectively and used as input.

The WSB+CAML pipeline was also run on the 11-taxon dataset having 100 genes and 500bp alignment length using binning threshold 75\% and compared with WSB+WQMC. The input gene trees for both WSB+WQMC and WSB+CAML were BestML gene trees estimated from gene alignments using RAxML version 7.3.5 with 200 bootstrap replicates. Supergene trees for each bin in the WSB-CAML pipeline were computed using BestML fully partitioned maximum likelihood analysis using RAxML version 8.2.7.
\subsubsection{SimPhy}\label{section:simphytesting}
The WSB+WQMC pipeline was run on the SimPhy datasets under various model conditions, as shown in Table \ref{table:testing}. For model conditions with 25, 50, and 100 genes, the first 25, 50, and 100 gene trees were sampled respectively and used as input. For model conditions with sequence lengths 50bp and 100bp, the first 50 and 100 sites were sampled from the original alignment. 

The input gene trees were estimated from the gene alignments using FastTree 2.1 without bootstrap values. In order to do the binning step in  the WSB+WQMC and WSB+CAML pipelines, the FastTree local support values were used instead of bootstrap support values. Apart from using the FastTree local support values instead of bootstrap support values, everything else was kept unchanged for running the WSB+WQMC and WSB+CAML pipelines. Based on the results obtained in the training phase, the binning threshold value of 95\% was chosen to be use in the testing phase.

The WSB+CAML pipeline was also run on the SimPhy datasets with model conditions having 100 genes and 100bp alignment length and compared with WSB+WQMC. Supergene trees for each bin in the WSB-CAML pipeline were computed using BestML unpartitioned maximum likelihood analysis using FastTree version 2.1.
\section{Methods}
\label{section:methods}
This section describes in detail the software and methods used for new gene tree and species tree estimation in this experimental study. 

\subsection{New gene tree estimation}
The WSB+WQMC pipeline computes a set of weighted quartets for each gene within a bin. WQMC is then run with the unique weighted quartet set as input to compute a new gene tree. Since the binning step computes disjoint bins, a new gene tree is obtained for each gene. The WSB+CAML pipeline on the other hand computes a supergene alignment for each bin. Maximum likelihood gene tree estimation method (RAxML or FastTree) is then run on each bin to compute a supergene tree for that bin. Finally, for each gene within a bin, the supergene tree computed for that bin is considered as the new gene tree.

\subsection{Species tree estimation}
To compute species trees from gene trees, summary methods ASTRID 1.1 \cite{vachaspati2015astrid} and ASTRAL2 4.7.2 \cite{mirarab2015astral} were used for the basic unbinned, WSB+WQMC, and WSB+CAML pipelines. 

ASTRAL2 in the default mode uses the bipartitions from the input gene trees to constraint the search space for the output species tree. To expand this search space, ASTRAL2 allows a set of extra bipartitions to be given as input. In this thesis, ASTRAL2 was run on three different sets of gene trees: the original set of gene trees, the newly estimated WSB+WQMC gene trees, and the newly estimated WSB+CAML gene trees.  To enable a fair comparison between these three ways of computing species trees, the same set of extra bipartitions for ASTRAL2 was used. This set of extra bipartitions contained bipartitions from all the trees listed below.
\begin{itemize}
    \item Original gene trees.
    \item New gene trees computed by running WSB+WQMC.
    \item Species tree computed by running ASTRID on the original gene trees.
    \item Species tree computed by running ASTRID on the new gene trees computed by WSB+WQMC.
    \item Species tree computed by running ASTRAL2 in default mode on the original gene trees.
\end{itemize}

\subsection{Measurement}
\label{section:measuremnt}

The new gene trees computed by both WSB+WQMC and WSB+CAML were evaluated by calculating gene tree estimation error. Gene tree estimation error  was computed by measuring the average missing branch rate (FN rate) of the new gene trees with respect to the true gene trees. Gene tree estimation error for the basic unbinned pipeline was calculated by measuring the average missing branch rate of the original gene trees with respect to the true gene trees.

Similarly, species tree estimation error was computed for both WSB+WQMC and WSB+CAML by measuring the average missing branch rate of the estimated species tree on new gene trees with respect to the true species tree. For the basic unbinned pipeline, species tree estimation error was calculated by measuring the average missing branch rate of the estimated species tree on original gene trees with respect to the true species tree.

For both gene tree estimation error and species tree estimation error, the average was computed over 10 replicates in each dataset.   
\newpage
\section{Figures and Tables}

\begin{table}[ht]
\centering
\begin{tabular}{llllllllll} 
\hline
Dataset &
Taxa & AD & Genes & Sites  & Reps & Bootstraps& Ref \\
\hline
Mammalian-0.5X & 37 & 50 & 50,200 &  500 & 10 & Available & \cite{mirarab2014statistical} \\
Mammalian-1X & 37 & 30 & 50,200 &  500 & 10 & Available & \cite{mirarab2014statistical} \\
Mammalian-2X & 37 & 21 & 50,200 &  500 & 10 & Available & \cite{mirarab2014statistical} \\
\hline
\end{tabular}
\caption{\textbf{Empirical statistics of simulated datasets used in training phase.} The ILS level is measured by AD\%, which is the percentage of missing branches between true gene trees and the species tree. Bootstraps column indicates whether bootstrap support values were available or not on estimated gene trees.}
\label{table:training}
\end{table}

\begin{table}[ht]
\centering
\begin{tabular}{llllllllll} 
\hline
Dataset &
Taxa & AD & Genes & Sites  & Reps & Bootstraps & Ref \\
\hline
Mammalian-1X & 37 & 30 & 50,200 &  250 & 10 & Available & \cite{mirarab2014statistical} \\
10-taxon H ILS & 10 & 84 & 25,50,100,200 & 100 & 10 & Available & \cite{bayzid2015weighted}\\ 
10-taxon MH ILS & 10 & 40 & 25,50,100,200 & 100 & 10 & Available & \cite{bayzid2015weighted}\\ 
15-taxon H ILS & 15 & 82 & 25,50,100,200,1000 & 100,1000 & 10 & Available & \cite{bayzid2015weighted}\\
11-taxon L ILS & 11 & 15 & 25,50,100 & 500 & 10 & Available & \cite{bayzid2013naive}\\
SimPhy-10M1e-7 & 50 & 9 & 25,50,100,200 & 50,100,200 & 10 & NA & \cite{mirarab2015astral}\\
SimPhy-10M1e-6 & 50 & 21 & 25,50,100,200 & 50,100,200 & 10 & NA & \cite{mirarab2015astral}\\
SimPhy-2M1e-6 & 50 & 30 & 25,50,100,200 & 50,100,200 & 10 & NA & \cite{mirarab2015astral}\\
SimPhy-2M1e-7 & 50 &34 & 25,50,100,200 & 50,100,200 & 10 & NA & \cite{mirarab2015astral}\\
SimPhy-500K1e-7 & 50 &68 & 100 & 100 & 10&NA & \cite{mirarab2015astral}\\
SimPhy-500K1e-6 & 50 &69 & 100 & 100 & 10 & NA & \cite{mirarab2015astral}\\
\hline
\end{tabular}
\caption{\textbf{Empirical statistics of simulated datasets used in testing phase.} The ILS level is measured by AD\%, which is the percentage of missing branches between true gene trees and the species tree. Bootstraps column indicates whether bootstrap support values were available or not on estimated gene trees.}
\label{table:testing}
\end{table}

\begin{table}[ht]
\centering
\begin{tabular}{cc} 
\hline
ILS Level & Range of AD\% \\
\hline
Low (L)& $< 22\%$ \\
Medium (M)& $22\% - 30\%$ \\
Moderately high (MH) & $31\% - 40\%$ \\
High (H) & $> 40\%$ \\
\hline
\end{tabular}
\caption{\textbf{Different categories of ILS level used in the study.} The ILS level is measured by AD\%, which is the percentage of missing branches between true gene trees and the species tree.}
\label{table:ils}
\end{table}

\begin{table}[ht]
\centering
\begin{tabular}{llcc} 
\hline
Genes & Sites & Avg bin-size ($t =90\%$)  & Avg bin-size ($t=95\%$)  \\
\hline
200 & 50 & 2.23 & 6.12 \\
100 & 50 & 2.09 & 5.42 \\
200 & 100 & 2.85 & 7.64 \\
100 & 100 & 2.55 & 6.93 \\
200 & 200 & 2.65 & 7.30 \\
100 & 200 & 2.47 & 6.29 \\

\hline
\end{tabular}
\caption{\textbf{Average bin-sizes for the SimPhy-10M1e-7 (9\% AD) dataset for different model conditions and binning thresholds $t$.} Statistical Binning was used to compute bins with different threshold values $t$. FastTree local support values were used instead  of  bootstrap support values for running Statistical Binning. The average was taken over 10 replicates.} 
\label{table:simphybins}
\end{table}

\end{doublespace}
\chapter{Methods and Commands}
\begin{doublespace}
\section{Data availability}
All datasets and supporting online materials are available at \url{	
http://goo.gl/zR48CT}. A github repository containing all the source code used in the experiments can be found at \url{https://github.com/agupta0905/improving_genes}.
\section{Commands}
\subsection*{Gene tree estimation}
RAxML 7.3.5 was used in \cite{bayzid2015weighted,mirarab2014statistical,bayzid2013naive} for the Mammalian, 10-taxon, 15-taxon and 11-taxon datasets to estimate each original gene tree from its sequence alignment with bootstrap support with the following commands.
\begin{verbatim}
    raxmlHPC-SSE3 -m GTRGAMMA -s [input_alignment] -n [output_name] -N 20
    -p [random_number]
\end{verbatim}
The following command was used for bootstrapping:
\begin{verbatim}
    raxmlHPC-SSE3 -m GTRGAMMA -s [input_alignment] -n [output_name] -N 200
    -p [random_number] -b [random_number]
\end{verbatim}
FastTree 2.1 was used to estimate each original gene tree from its sequence alignment for the SimPhy dataset using the following command.
\begin{verbatim}
    FastTree -nt -gtr -gamma < [input_alignment] > [output_genetree]
\end{verbatim}
WQMC was used to estimate each new gene tree from its weighted quartet file in the WSB+WQMC pipeline using the following command.
\begin{verbatim}
   max-cut-tree qrtt=[weight_quartet_file] weights=on otre=[output_new_genetree]
\end{verbatim}
\subsection*{Supergene tree estimation}
RAxML 8.2.7 was used to estimate the supergene tree from the supergene alignment for the 10-taxon, 15-taxon and 11-taxon datasets in fully partitioned mode with the following command.
\begin{verbatim}
    raxmlHPC-SSE3 -m GTRGAMMA -s [input_alignment] -n [output_name] -N 20 
    -M -q [partition_file] -p [random_number]
\end{verbatim}
RAxML 8.2.7 was used to estimate the supergene tree from the supergene alignment for the mammalian dataset in unpartitioned mode with the following command.
\begin{verbatim}
    raxmlHPC-SSE3 -m GTRGAMMA -s [input_alignment] -n [output_name] -N 20 
    -p [random_number]
\end{verbatim}
FastTree 2.1 was used to estimate the supergene tree from the supergene alignment for the SimPhy dataset in unpartitioned mode with the following command.
\begin{verbatim}
    FastTree -nt -gtr -gamma < [input_alignment] > [output_genetree]
\end{verbatim}
\subsection*{Species tree estimation}
ASTRID 1.1 was used to estimate the species tree from original gene trees, WSB+WQMC gene trees, and WSB+CAML gene trees with the following command.
\begin{verbatim}
    ASTRID -i [input_gene_trees] -o [output_species_tree]
\end{verbatim}
ASTRAL2 4.7.2 was used in default mode to estimate the species tree from original gene trees with the following command.
\begin{verbatim}
    java -jar astral.4.7.12.jar -i [input_gene_trees] -o [output_species_tree]
\end{verbatim}
ASTRAL2 4.7.2 was used with a set of extra bipartitions to estimate the species tree from original gene trees, WSB+WQMC gene trees, and WSB+CAML gene trees with the following command.
\begin{verbatim}
    java -jar astral.4.7.12.jar -i [input_gene_trees] -o [output_species_tree] 
    -e [extra_gene_trees]
\end{verbatim}
\end{doublespace}
\chapter{Results}
\label{chapter:results}
\begin{doublespace}
\section{Training}
\subsection*{Impact of confidence value c and different binning thresholds on gene tree estimation error}
\begin{figure}[ht]
\centering
\includegraphics[width=\textwidth]{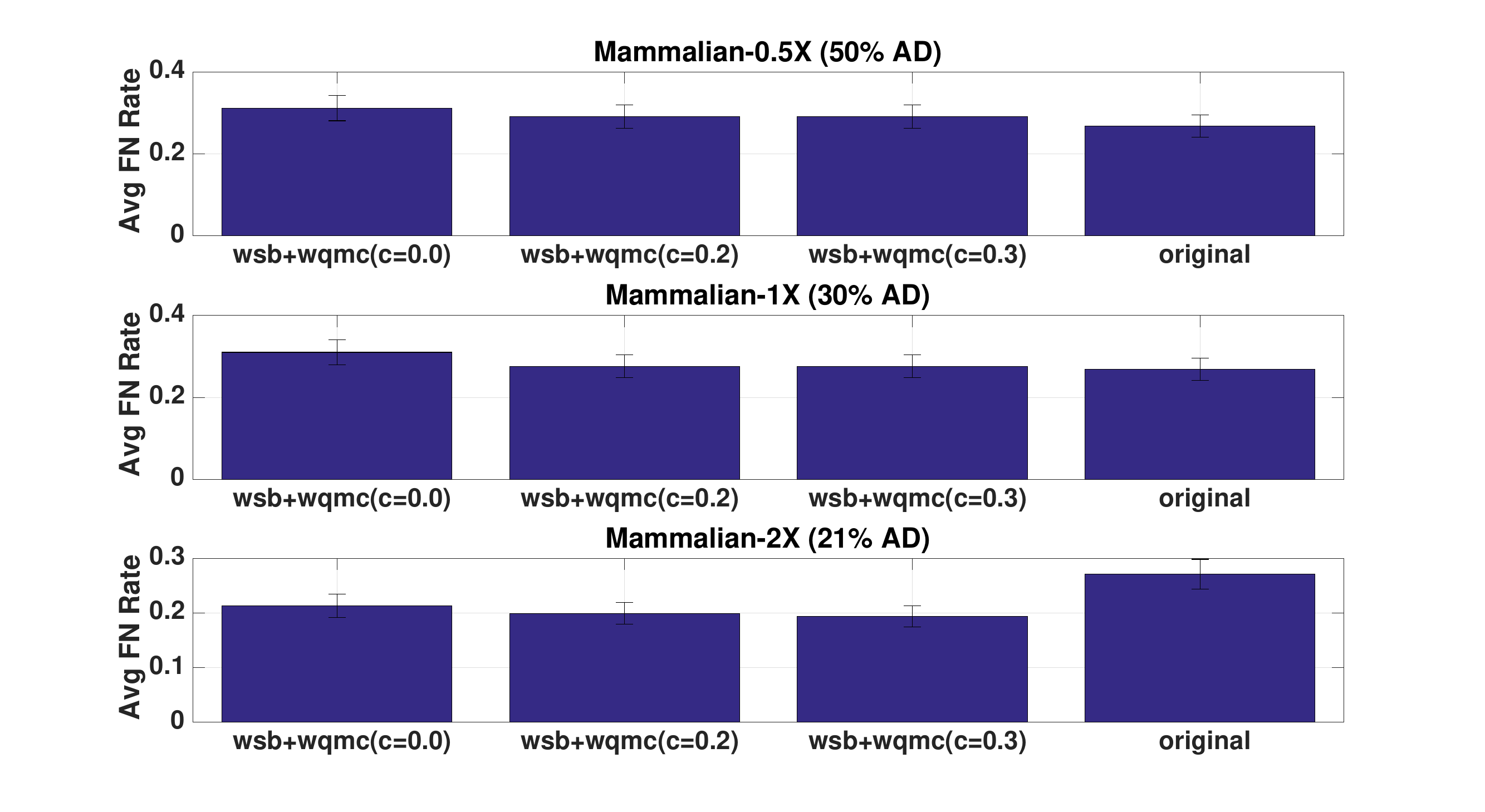}
\caption{\textbf{Gene tree estimation error on the Mammalian datasets for running WSB+WQMC with different confidence values and binning threshold 75\%.}  We show results on the Mammalian-0.5X (50\% AD), Mammalian-1X (30\% AD), and Mammalian-2X (21\% AD) datasets with 200 genes and 500bp alignment length. The x-axis shows three different ways of running WSB+WQMC by varying confidence value c (0.0, 0.2, and 0.3). The accuracy of WSB+WQMC gene trees is also compared to the accuracy of original gene trees. Average FN rate is shown with standard error bars over 10 replicates.}
\label{fig:training-gene}
\end{figure}

Gene tree estimation results on the Mammalian-0.5X, Mammalian-1X, and Mammalian-2X datasets with 200 genes and 500bp alignment length by running the WSB+WQMC pipeline with varying confidence values and binning threshold 75\% are shown in Figure \ref{fig:training-gene}. It was observed that the new gene trees computed by WSB+WQMC with confidence value 0.0 were consistently worse than WSB+WQMC with confidence values 0.2 and 0.3 among all ILS levels. WSB+WQMC with confidence value 0.2 and 0.3 produced similar results and were very close to each other. Compared to the original gene trees, the WSB+WQMC pipeline with confidence value 0.2 and 0.3 produced worse gene trees on high ILS Mammalian-0.5X (50\% AD), similar gene trees on medium ILS Mammalian-1X (30\% AD), and better gene trees on low ILS Mammalian-0.5X (21\% AD). 

\begin{figure}[p]
\centering
\includegraphics[width=\textwidth]{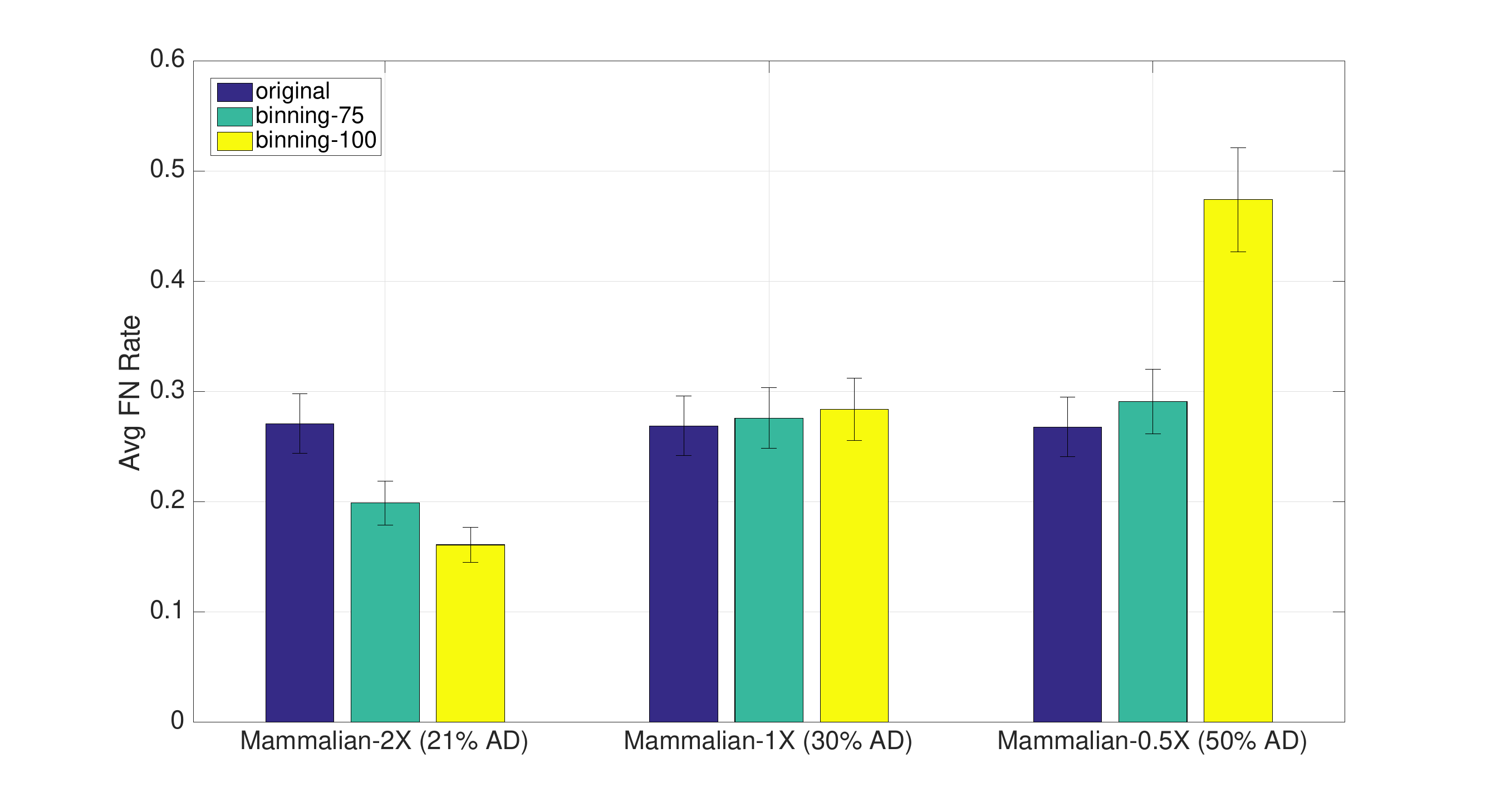}
\caption{\textbf{Gene tree estimation error on the Mammalian datasets for running WSB+WQMC with binning threshold 75\% and binning threshold 100\% using confidence value 0.2.} We show results on the Mammalian-0.5X (50\% AD), Mammalian-1X (30\% AD), and Mammalian-2X (21\% AD) datasets with 200 genes and 500bp alignment length. The accuracy of WSB+WQMC gene trees is also compared to the accuracy of original gene trees. Average FN rate is shown with standard error bars over 10 replicates.}
\label{fig:training-binning-gene}
\end{figure}

Figure \ref{fig:training-binning-gene} shows gene tree estimation results on the Mammalian-0.5X, Mammalian-1X, and Mammalian-2X datasets with 200 genes and 500bp alignment length by running the WSB+WQMC pipeline using binning threshold 75\% and binning threshold 100\% (one-bin) with confidence value 0.2. It was observed that binning 75\% outperformed binning 100\% in medium and high ILS conditions, with binning 100\% being much worse and computing highly inaccurate gene trees for high ILS. In low ILS conditions, binning 100\% performed better than binning 75\%. Compared to the original gene trees, WSB+WQMC computed better gene trees in low ILS conditions, slightly worse gene trees in medium ILS, and worse gene trees in high ILS conditions. 

We observed that the new gene trees got better on the Mammalian-2X dataset (21\% AD), were similar to the original gene trees on the Mammalian-1X dataset (30\% AD), and got worse in the Mammalian-0.5X dataset (50\% AD) for both binning thresholds. Also, it was observed that between binning 100\% and binning 75\%, binning 100\% got higher magnitudes of change with respect to the original gene trees (substantially worse in high ILS and more accurate in low ILS).

\subsection*{Impact of confidence value c and different binning thresholds on species tree estimation error}

\begin{figure}[p]
\centering
\includegraphics[width=\textwidth]{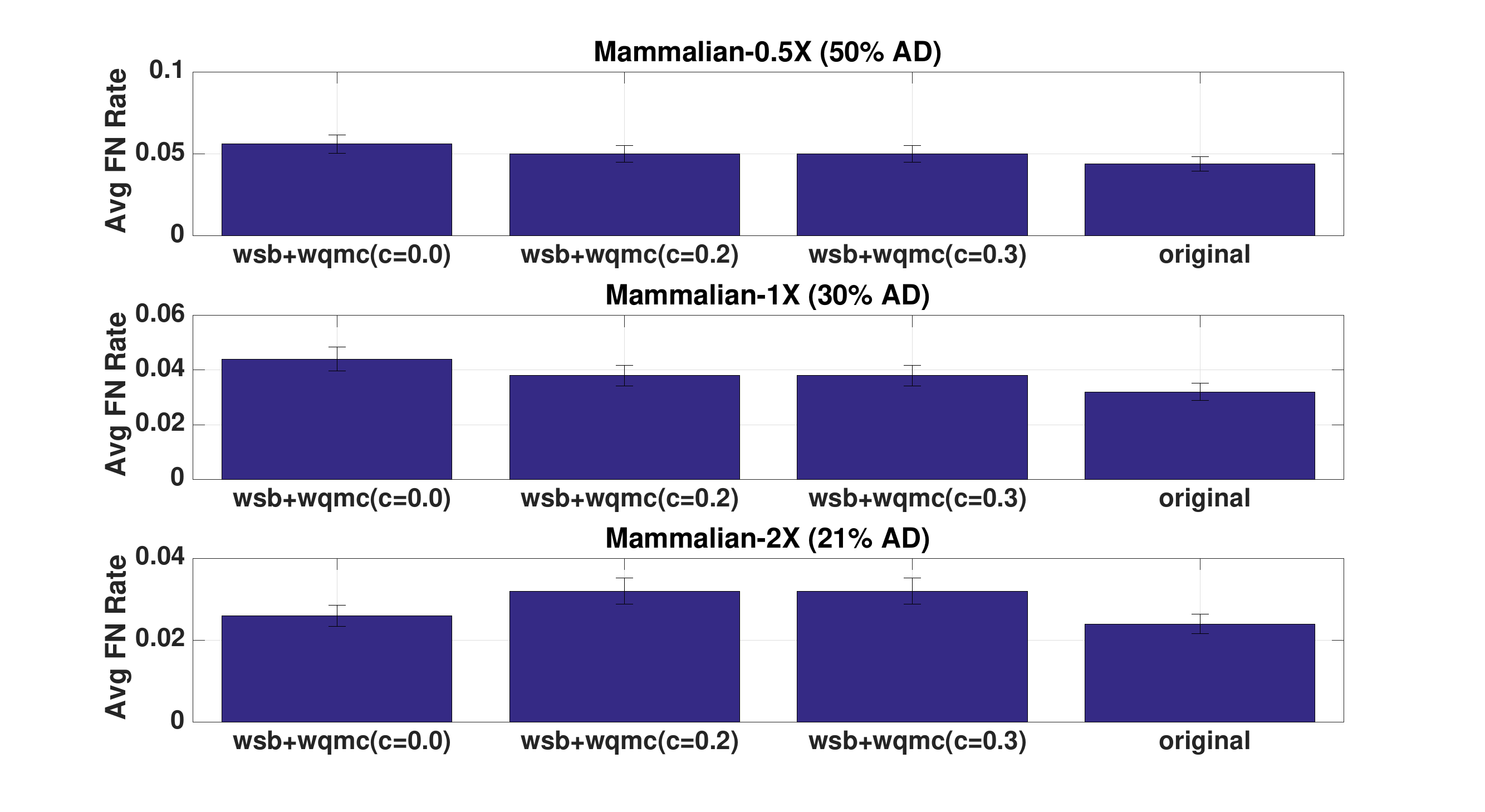}
\caption{\textbf{ASTRAL2 species tree estimation error on the Mammalian datasets for running WSB+WQMC with different confidence values and binning threshold 75\%.}  We show results on the Mammalian-0.5X (50\% AD), Mammalian-1X (30\% AD), and Mammalian-2X (21\% AD) datasets with 200 genes and 500bp alignment length. The x-axis shows three different ways of running WSB+WQMC by varying confidence value c (0.0, 0.2, and 0.3). The accuracy of species tree on WSB+WQMC gene trees is also compared to the accuracy of species tree on original gene trees. Average FN rate is shown with standard error bars over 10 replicates. }
\label{fig:training-astral2}
\end{figure}

ASTRAL2 species tree estimation results on the Mammalian-0.5X, Mammalian-1X, and Mammalian-2X datasets with 200 genes and 500bp alignment by running the WSB+WQMC pipeline with varying confidence values and binning threshold 75\% are shown in Figure \ref{fig:training-astral2}. It was observed that WSB+WQMC with confidence value 0.0 was worse than confidence values 0.2 and 0.3 for the Mammalian-0.5X (50\% AD) and Mammalian-1X (30\% AD) datasets but better for the Mammalian-2X (21\% AD) dataset. Compared to each other, confidence values 0.2 and 0.3 were very close to each other. For all ILS levels and confidence values, it was observed that WSB+WQMC failed to compute a better species tree than the species tree on original gene trees. 

\begin{figure}[p]
\centering
\includegraphics[width=\textwidth]{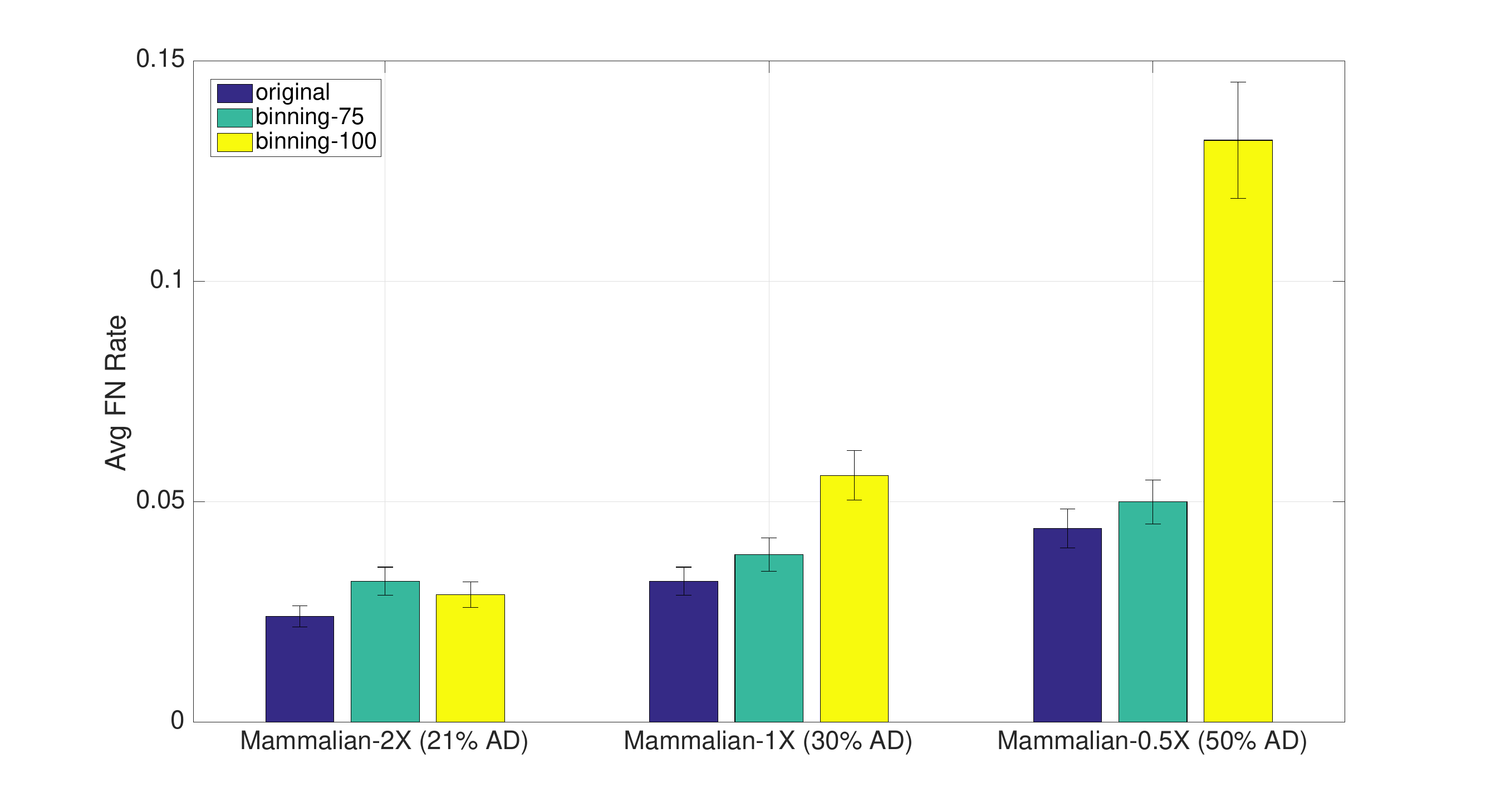}
\caption{\textbf{ASTRAL2 species tree estimation error on the Mammalian datasets for running WSB+WQMC with binning threshold 75\% and binning threshold 100\% using confidence value 0.2.} We show results on the Mammalian-0.5X (50\% AD), Mammalian-1X (30\% AD), and Mammalian-2X (21\% AD) datasets with 200 genes and 500bp alignment length. The accuracy of species tree on WSB+WQMC gene trees is also compared to the accuracy of species tree on original gene trees. Average FN rate is shown with standard error bars over 10 replicates.}
\label{fig:training-binning-astral2}
\end{figure}

Figure \ref{fig:training-binning-astral2} shows ASTRAL2 species tree estimation results on the Mammalian-0.5X, Mammalian-1X, and Mammalian-2X datasets with 200 genes and 500bp alignment length by running the WSB+WQMC pipeline using binning threshold 75\% and binning threshold 100\% (one-bin) with confidence value 0.2. It was observed that binning 75\% outperformed binning 100\% in medium and high ILS conditions, with binning 100\% being much worse and computing highly inaccurate species trees in high ILS conditions. In low ILS conditions, binning 100\% performed better than binning 75\%. Overall, both binning thresholds used for running  WSB+WQMC failed to compute a more accurate species tree compared to the species tree computed on  the original gene trees.

\section{Testing}
\subsection*{Impact of WSB+WQMC on gene tree estimation error}
In the following section, the accuracy of new gene trees computed by running the WSB+WQMC pipeline is compared to the accuracy of original gene trees.  

Gene tree estimation results after running WSB+WQMC on all datasets and all model conditions listed in Tables \ref{table:testing} and \ref{table:training}, with an exception of the SimPhy-500K1e-7 and SimPhy-500K1e-6 datasets, are shown in Figure \ref{fig:testing-gene}.
In general, in low ILS levels the WSB+WQMC gene trees got more accurate than the original gene trees. With increasing ILS, the improvement in the accuracy of WSB+WQMC gene trees over the original gene trees decreased and finally WSB+WQMC gene trees got less accurate in  high ILS conditions. Results show that WSB+WQMC improves gene tree accuracy for datasets  having low level of ILS and decreases accuracy of gene trees for datasets having high levels of ILS. For medium and moderately high ILS levels, the relative performance of WSB+WQMC compared to  the basic unbinned pipeline was dependent on the dataset and model condition itself. The Mammalian-1X  (30\% AD) and SimPhy 2M1e-7 (34\% AD) datasets were better off using the WSB+WQMC pipeline on some model conditions but worse for others. On the other hand, on the 10-taxon MH ILS dataset (40\% AD) it was better to use WSB+WQMC than the basic unbinned pipeline on all model conditions.  

\begin{figure}[p]
\centering
\includegraphics[width=\textwidth]{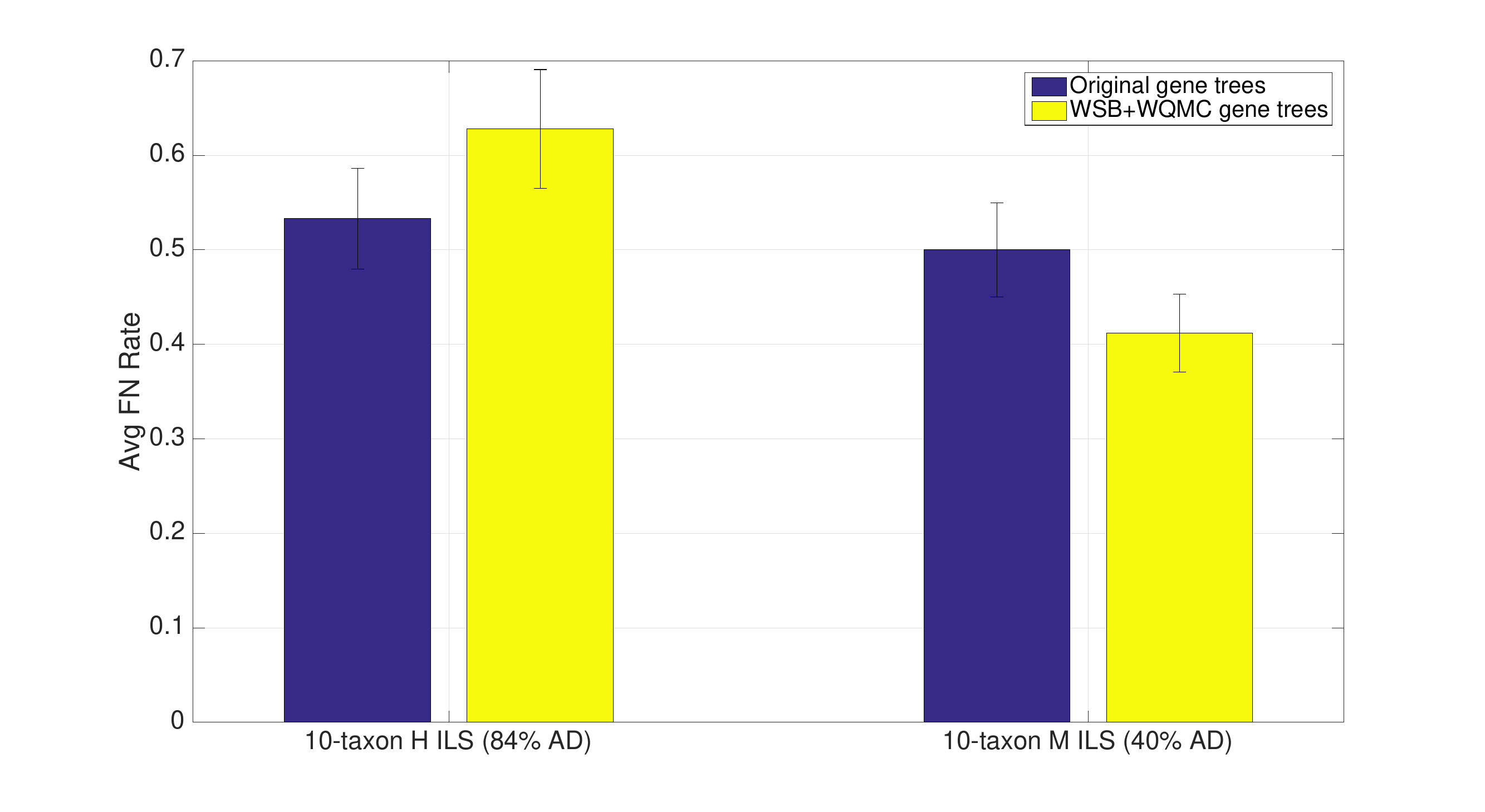}
\caption{\textbf{Gene tree estimation error on 10-taxon datasets for running WSB+WQMC.} We show results on the 10-taxon MH ILS (40\% AD) and 10-taxon H ILS (84\% AD) with 200 genes and 100bp alignment length. WSB+WQMC was run using binning threshold 75\% and confidence value 0.2. The accuracy of WSB+WQMC gene trees is also compared to the accuracy of original gene trees. Average FN rate is shown with standard error bars over 10 replicates.}
\label{fig:testing-10taxon-gene}
\end{figure}

\begin{figure}[p]
\centering
\includegraphics[width=\textwidth]{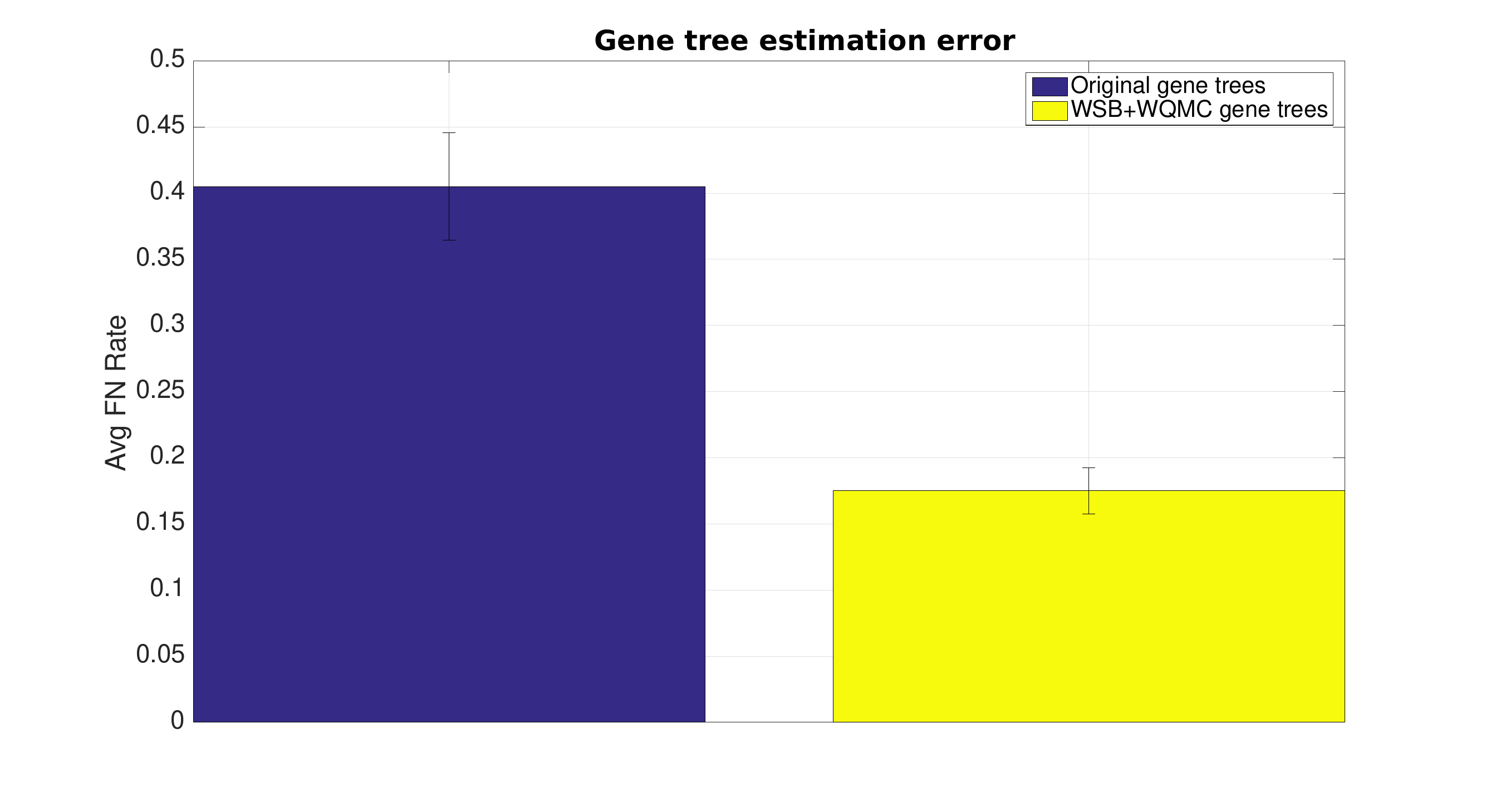}
\caption{\textbf{Gene tree estimation error on 11-taxon dataset for running WSB+WQMC.} We show results on the 11-taxon L ILS (15\% AD) dataset was used having 100 genes and 500bp alignment length. WSB+WQMC was run using binning threshold 75\% and confidence value 0.2. The accuracy of WSB+WQMC gene trees is also compared to the accuracy of original gene trees. Average FN rate is shown with standard error bars over 10 replicates.}
\label{fig:testing-11taxon-gene}
\end{figure}

\begin{figure}[p]
\centering
\includegraphics[width=\textwidth]{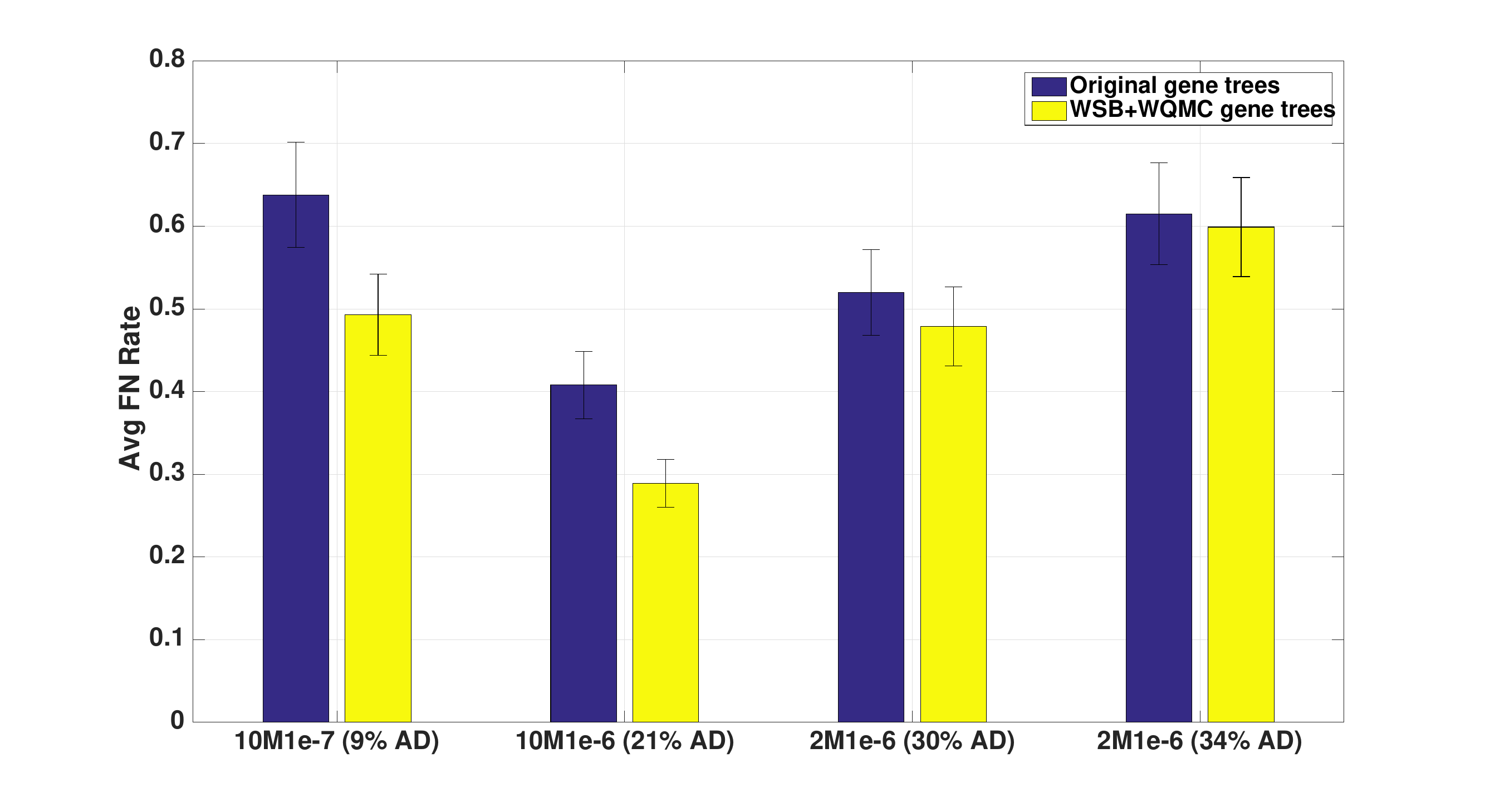}
\caption{\textbf{Gene tree estimation error on SimPhy datasets for running WSB+WQMC.} We show results on the 10M1e-7 (9\% AD), 10M1e-6 (21\% AD), 2M1e-6 (30\% AD), and 2M1e-7 (34\% AD)  datasets with 200 genes and 200bp alignment length. WSB+WQMC was run using binning threshold 95\% and confidence value 0.2. The accuracy of WSB+WQMC gene trees is also compared to the accuracy of original gene trees. Average FN rate is shown with standard error bars over 10 replicates.}
\label{fig:testing-simphy-gene}
\end{figure}

This strong relationship between the ILS level and the improvement in gene tree estimation error is also seen in Figure \ref{fig:testing-10taxon-gene}, where WSB+WQMC substantially improves the accuracy of gene trees for the 10-taxon MH ILS dataset (40\% AD, 200 genes and 100bp) but makes the gene trees worse for the 10-taxon H ILS dataset (84\% AD, 200 genes and 100bp). The degree of improvement in the accuracy of gene trees was also seen to be directly related to the ILS level. Results on the 11-taxon L ILS dataset shown in Figure \ref{fig:testing-11taxon-gene} and SimPhy datasets (except 500K1e-6 and 500K1e-7) shown in Figure \ref{fig:testing-simphy-gene} clearly show huge improvements in gene tree accuracy for the 11-taxon L ILS (15\% AD), 10M1e-7 (9\% AD), and 10M1e-6 datasets (21\% AD), whereas the improvements in gene tree accuracy for 2M1e-6 (30\% AD) and 2M1e-7 (34 \% AD) datasets having medium and moderately high ILS levels were lower in magnitude.    

\begin{figure}[p]
\centering
\includegraphics[width=\textwidth]{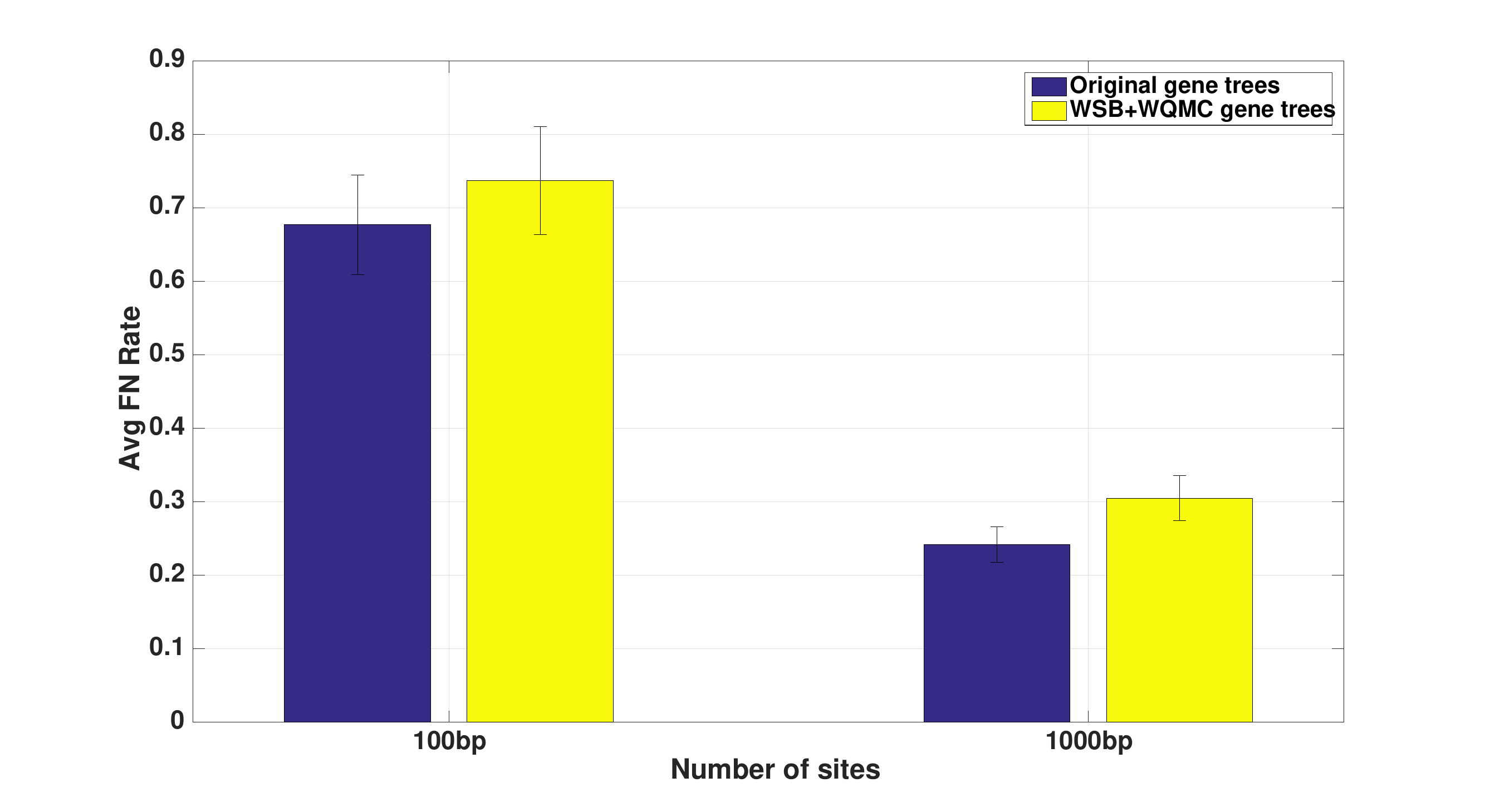}
\caption{\textbf{Gene tree estimation error on 15-taxon H ILS dataset with two different alignment lengths for running WSB+WQMC.} We show results on the 15-taxon H ILS (82\% AD) dataset was used having 1000 genes and two different alignment lengths (100bp and 1000bp). WSB+WQMC was run using binning threshold 75\% and confidence value 0.2. The accuracy of WSB+WQMC gene trees is also compared to the accuracy of original gene trees. Average FN rate is shown with standard error bars over 10 replicates. }
\label{fig:testing-15taxon-gene}
\end{figure}

The effect of varying alignment length on the accuracy of WSB+WQMC gene trees was also studied for the Mammalian-1X and 15-taxon H ILS datasets, as shown in Figures \ref{fig:training-numsites-gene} and \ref{fig:testing-15taxon-gene}. WSB+WQMC was run on the Mammalian-1X dataset (30\% AD) having 200 genes with two model conditions having 250bp and 500bp alignment lengths. For the 15-taxon H ILS dataset (82\% AD), WSB+WQMC was run on two different model conditions having 100bp and 1000bp alignment lengths. On the Mammalian-1X dataset, shorter alignment length improved the accuracy of gene trees substantially while longer alignment length computed worse gene trees compared to the original gene trees. For the 15-taxon H ILS dataset both alignment lengths computed better WSB+WQMC gene trees than the original gene trees. However, in the model condition having 1000bp alignment length, the accuracy decreased by a greater factor compared to the 100bp alignment length model condition. These results suggest that the benefit of using WSB+WQMC to compute gene trees compared to using the basic unbinned pipeline was greater for model conditions having smaller alignment lengths.      

\subsection*{Comparison of WSB+WQMC and WSB+CAML in terms of gene tree estimation error}

When comparing WSB+WQMC with WSB+CAML on the 11-taxon L ILS (15\% AD), Mammalian-1X (30\% AD), 10-taxon MH ILS (40\% AD), and 15-taxon H ILS (82\% AD) datasets (as shown in Figure \ref{fig:caml-others-gene}), WSB+WQMC computed better gene trees than WSB+CAML for all datasets and all ILS levels in the experiment. Both pipelines computed better gene trees than the original gene trees for the 11-taxon L ILS, Mammalian-1X, and 10-taxon MH ILS datasets but worse gene trees for the 15-taxon H ILS dataset. 

\begin{figure}[p]
\centering
\includegraphics[width=\textwidth]{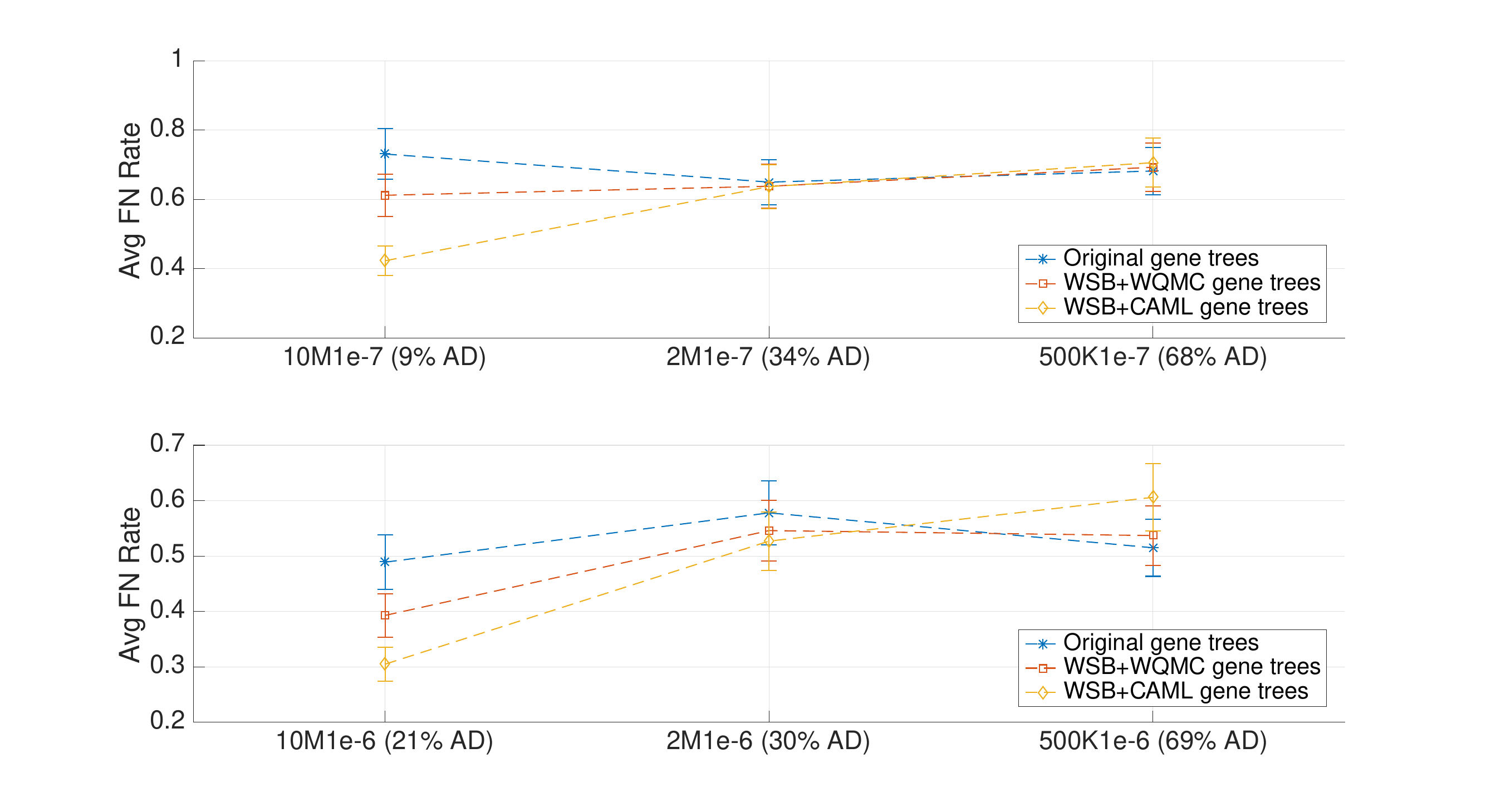}
\caption{\textbf{Gene tree estimation error on SimPhy datasets having speciation rate 1e-7 (top) and having speciation rate 1e-6 (bottom) for running WSB+WQMC and WSB+CAML.} We show results on the 10M1e-7 (9\% AD), 2M1e-7 (34\% AD), 500KM1e-7 (68\% AD), 10M1e-6 (21\% AD), 2M1e-6 (30\% AD), and 500K1e-6 (69\% AD) datasets with 100 genes and 100bp alignment length. WSB+WQMC was run using binning threshold 95\% and confidence value 0.2. WSB+CAML was run using binning threshold 95\%. The accuracy of WSB+WQMC and WSB+CAML gene trees is also compared to the accuracy of original gene trees. Average FN rate is shown with standard error bars over 10 replicates.}
\label{fig:caml-simphy-gene}
\end{figure}

In contrast, when comparing WSB+WQMC with WSB+CAML on the SimPhy datasets having speciation rates 1e-7 and 1e-6 (shown in Figure \ref{fig:caml-simphy-gene}), WSB+CAML computed better gene trees for low ILS and medium ILS datasets (i.e., 10M1e-6, 10M1e-7, and 2M1e-6) and worse gene trees for high ILS datasets (i.e., 500K1e-6 and 500K1e-7). For moderately high ILS datasets (i.e., 2M1e-7), both WSB+WQMC and WSB+CAML performed equally well in estimating gene trees. Compared to the original gene trees, both pipelines computed better gene trees in low, medium, and moderately high ILS levels. In high ILS levels, both pipelines computed worse gene trees compared to the original gene trees.    

In terms of gene tree estimation, we didn't find a clear winner between WSB+CAML and WSB-WQMC, as their relative performance was dependent on the ILS level and the dataset itself. It was found that WSB+WQMC computed better gene trees than WSB+CAML in high ILS conditions. In moderately high ILS levels, WSB+WQMC performed better than WSB+CAML on the 10-taxon MH ILS (40\% AD) dataset but similar to WSB+CAML on the SimPhy 2M1e-7 (34\% AD) dataset. In medium and low ILS conditions, WSB+WQMC computed better gene trees than WSB+CAML on the 11-taxon L ILS (15\% AD) and Mammalian-1X (30\% AD) but worse gene trees on the SimPhy 10M1e-7 (9\% AD), 10M1e-7(21\% AD), and 2M1e-6 (30\% AD) datasets.   

\subsection*{Impact of WSB+WQMC on species tree estimation error}
 In the following section, the accuracy of species trees computed by running the WSB+WQMC pipeline is compared to the accuracy of species trees computed by running the basic unbinned pipeline.  
 
 ASTRAL2 species tree estimation results on all datasets and all model conditions listed in Tables \ref{table:testing} and \ref{table:training}, with an exception of the SimPhy-500K1e-7 and SimPhy-500K1e-6 datasets, are shown in Figure \ref{fig:testing-astral2}. In general, in low ILS levels the ASTRAL2 species tree on WSB+WQMC gene trees got more accurate than the ASTRAL2 species tree on original gene trees. With increasing ILS, the improvement in accuracy of ASTRAL2 species tree using the WSB+WQMC pipeline decreased and finally the ASTRAL2 species tree computed on WSB+WQMC gene trees got less accurate in high ILS conditions. Although WSB+WQMC computed more accurate ASTRAL2 species trees on most model conditions having low ILS, there were a few exceptions. There was no clear pattern among these low ILS model conditions where WSB+WQMC was computing less accurate ASTRAL2 species trees. In medium and moderately high ILS levels, most model conditions benefited by using the WSB+WQMC pipeline, but there were some model conditions among these datasets where WSB+WQMC performed poorly. The results showed another interesting trend with the improvement in accuracy of ASTRAL2 species tree using the WSB+WQMC pipeline for some model conditions on the 15-taxon H ILS (82\% AD) and 10-taxon (84\% AD) datasets having high levels of ILS. Again, there was no clear pattern among the model conditions having high ILS and more accurate ASTRAL2 species trees computed using the WSB+WQMC pipeline.

\begin{figure}[p]
\centering
\includegraphics[width=\textwidth]{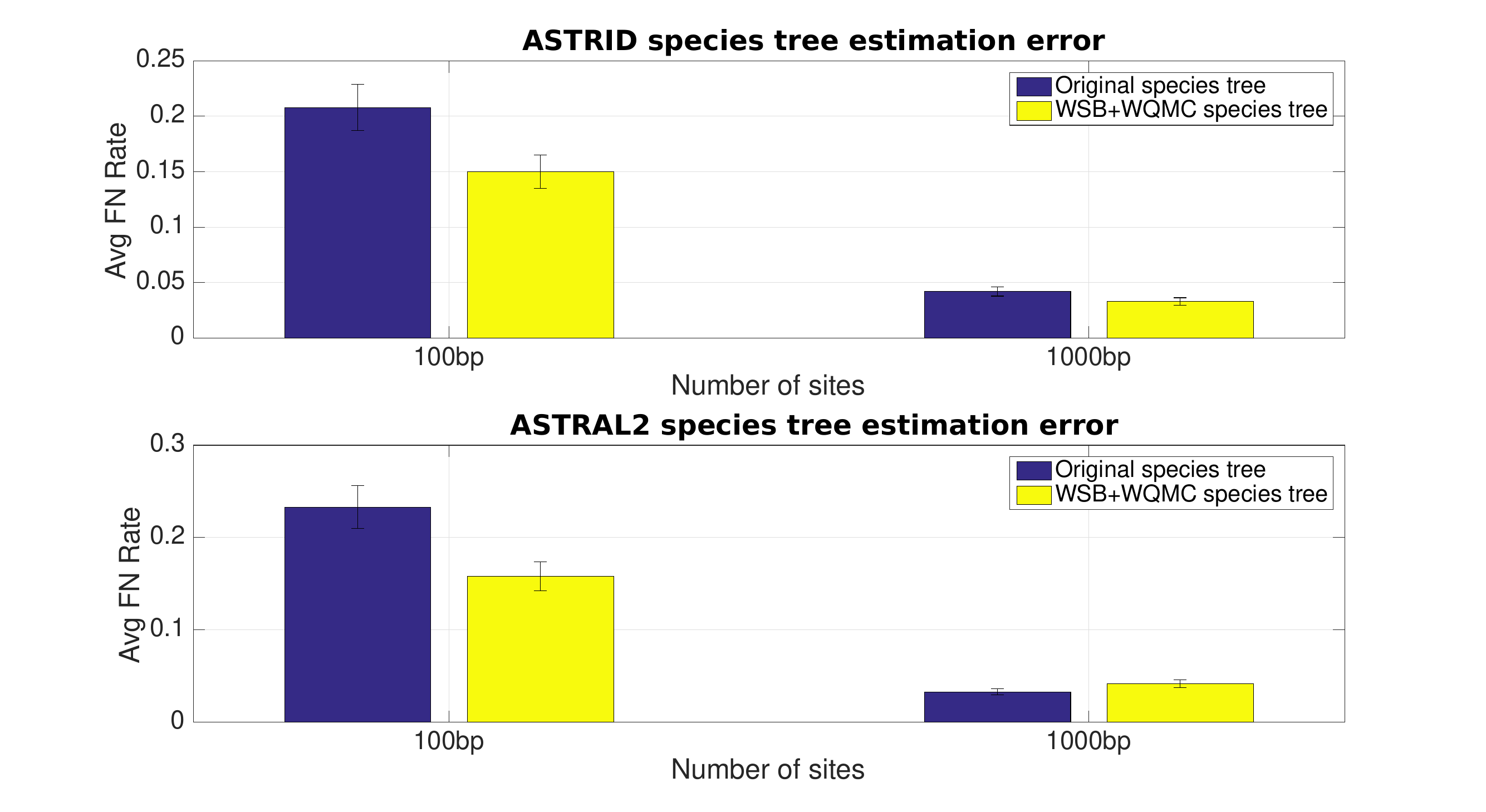}
\caption{\textbf{Species tree estimation error using ASTRID (top) and ASTRAL2 (bottom) on 15-taxon H ILS dataset with two different alignment lengths for running WSB+WQMC.} We show results on the 15-taxon H ILS (82\% AD) dataset was used having 1000 genes and two different alignment lengths (100bp and 1000bp). WSB+WQMC was run using binning threshold 75\% and confidence value 0.2. The accuracy of species tree on WSB+WQMC gene trees is also compared to the accuracy of species tree on original gene trees. Average FN rate is shown with standard error bars over 10 replicates. }
\label{fig:testing-15taxon-species}
\end{figure}

This weak correlation between the ILS level and the improvement in species tree estimation error using the WSB+WQMC pipeline was found throughout the experiments. Figure \ref{fig:testing-11taxon-species} shows results on the 11-taxon L ILS (15\% AD) dataset, where WSB+WQMC improved the accuracy of the ASTRID species tree but had no change for the ASTRAL2 species tree. ASTRID and ASTRAL2 species trees computed using WSB+WQMC gene trees on low, medium, and moderately high ILS SimPhy datasets (i.e., 10M1e-7, 10M1e-6, 2M1e-6, and 2M1e-7)  were seen to be substantially better than the species trees computed on original gene trees (see Figure \ref{fig:testing-simphy-species}). Results on the 10-taxon MH ILS (40\% AD) and 10-taxon H ILS (82\%) datasets (shown in Figure \ref{fig:testing-10taxon-species}) show that ASTRID and ASTRAL2 species trees computed on WSB+WQMC gene trees were less accurate than the species trees computed on original gene trees. In contrast, WSB+WQMC was able to reduce species tree estimation error for the 15-taxon H ILS (82\% AD) dataset (shown in Figure \ref{fig:testing-15taxon-species}), despite having high ILS. Both ASTRID and ASTRAL2 species trees got more accurate for the 15-taxon dataset with 100bp alignment length model condition but only the ASTRID species tree was improved for the 15-taxon 1000bp model condition. When compared to the unbinned pipeline, it was also observed that on the 15-taxon dataset WSB+WQMC performed better in the shorter alignment model condition (100bp) than the longer alignment model condition (1000bp), with greater magnitude of improvements in the former case. Figure \ref{fig:training-numsites-gene} shows the impact of alignment length on the accuracy of species trees computed by WSB+WQMC on the Mammalian-1X (30\% AD) dataset. In this experiment, alignment lengths of 250bp and 500bp were used. It was found that for both model conditions, ASTRAL2 and ASTRID species trees on WSB+WQMC gene trees were worse than the species trees on original gene trees. Varying alignment length didn't seem to affect the performance of WSB+WQMC on this dataset. 

\subsection*{Comparison of WSB+WQMC and WSB+CAML in terms of species tree estimation error}

\begin{figure}[p]
\centering
\includegraphics[width=\textwidth]{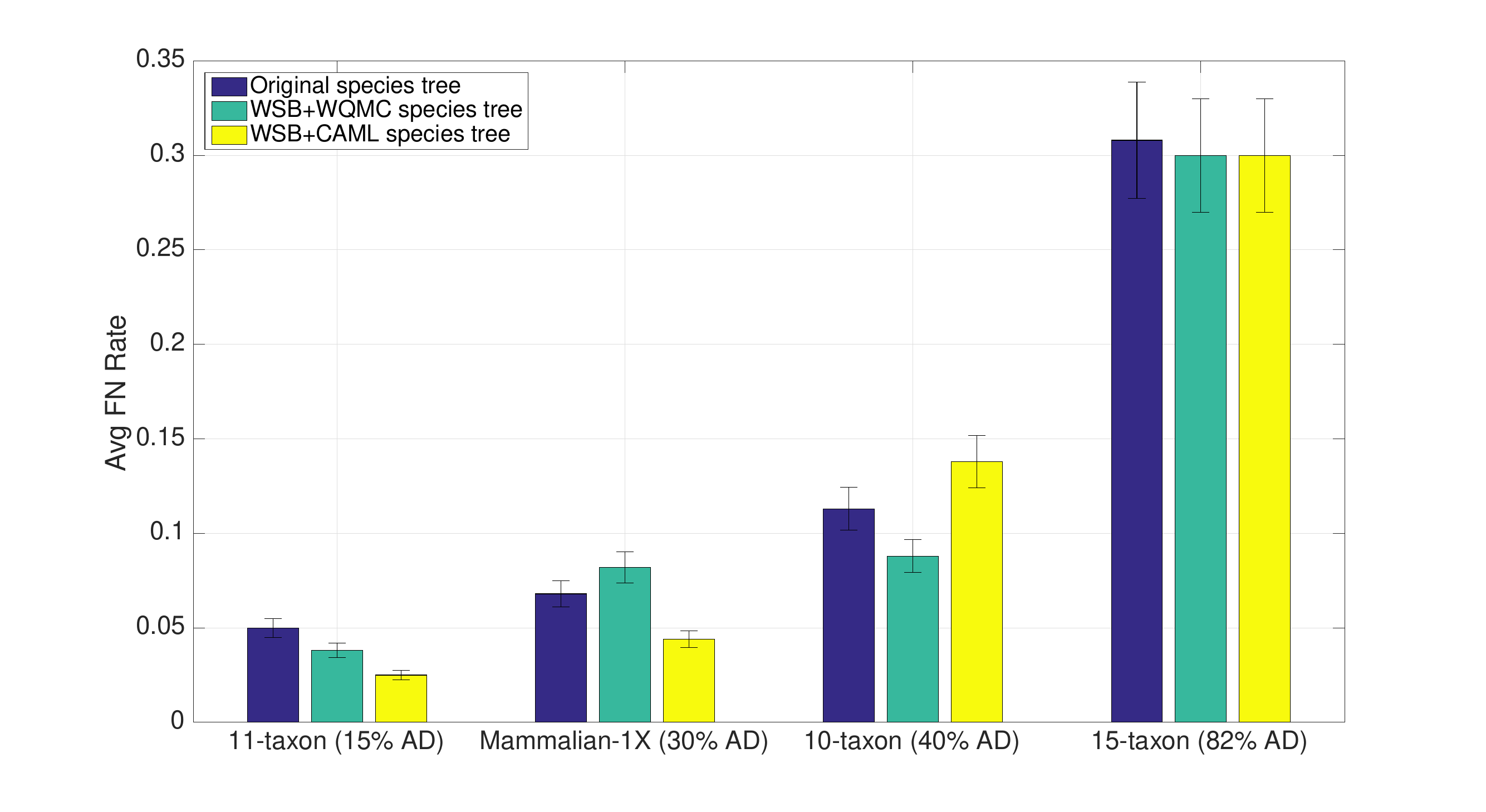}
\caption{\textbf{ASTRID species tree estimation error on 11-taxon, Mammalian-1X, 10-taxon and 15-taxon datasets for running WSB+WQMC and WSB+CAML.} We show results on the 11-taxon L ILS (100 genes and 500bp),  Mammalian-1X (200 genes, 250bp), 10-taxon MH ILS (100 genes, 100bp), and 15-taxon H ILS (100 genes and 100bp) datasets were used. The level of ILS increases from left to right. WSB+WQMC was run using binning threshold 75\% and confidence value 0.2. WSB+CAML was run using binning threshold 75\%. The accuracy of species tree on WSB+WQMC and WSB+CAML gene trees were also compared to the accuracy of species tree on original gene trees. Average FN rate is shown with standard error bars over 10 replicates.}
\label{fig:caml-others-astrid}
\end{figure}

\begin{figure}[p]
\centering
\includegraphics[width=\textwidth]{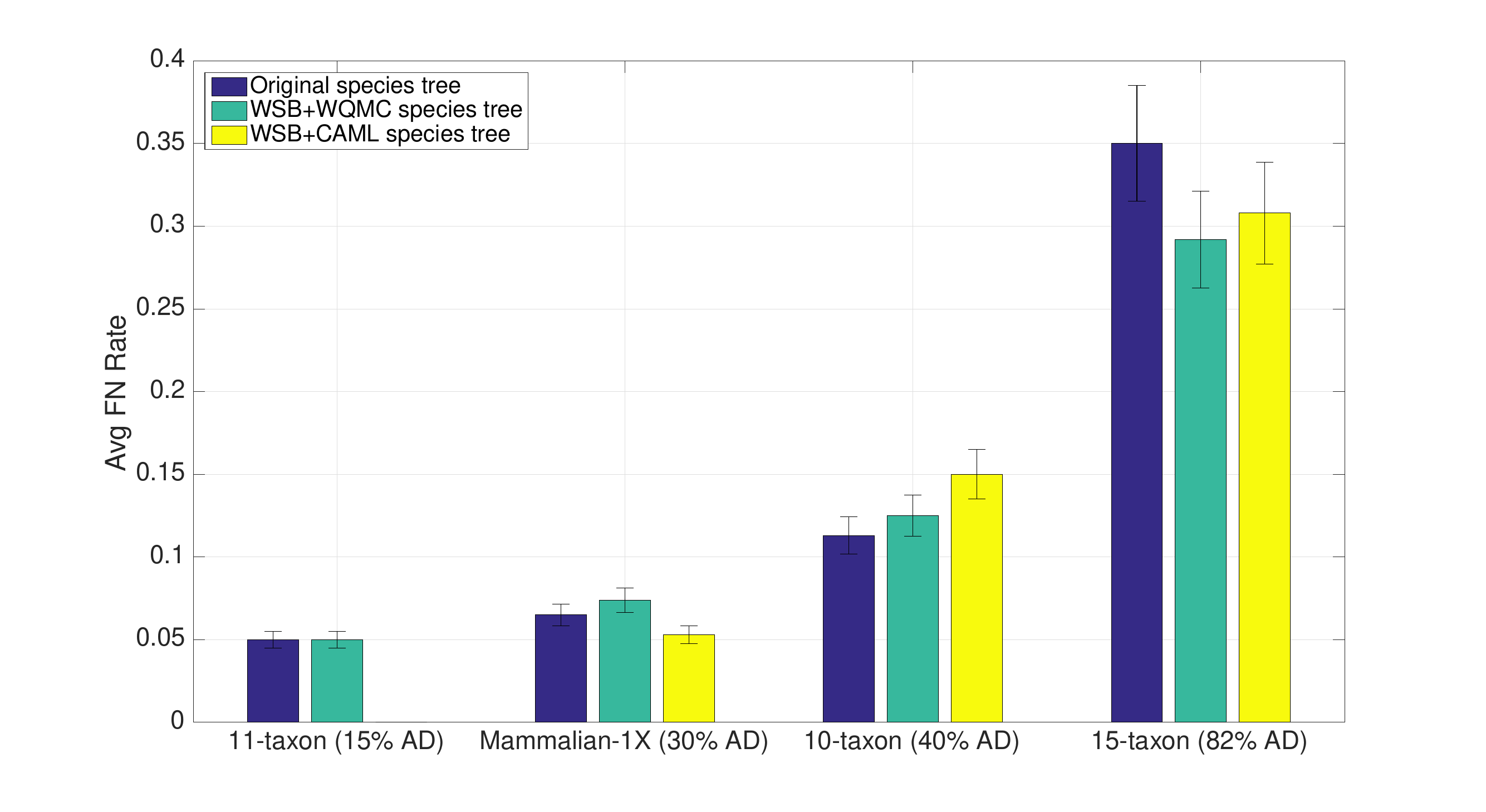}
\caption{\textbf{ASTRAL2 species tree estimation error on 11-taxon, Mammalian-1X, 10-taxon, and 15-taxon datasets for running WSB+WQMC and WSB+CAML.} We show results on the 11-taxon L ILS (100 genes and 500bp),  Mammalian-1X (200 genes, 250bp), 10-taxon MH ILS (100 genes, 100bp) and 15-taxon H ILS (100 genes and 100bp) datasets were used. The level of ILS increases from left to right. WSB+WQMC was run using binning threshold 75\% and confidence value 0.2. WSB+CAML was run using binning threshold 75\%. The accuracy of species tree on WSB+WQMC and WSB+CAML gene trees were also compared to the accuracy of species tree on original gene trees. Average FN rate is shown with standard error bars over 10 replicates.}
\label{fig:caml-others-astral2}
\end{figure}

  Figures \ref{fig:caml-others-astrid} and \ref{fig:caml-others-astral2} show the species tree estimation results by running the WSB+WQMC and WSB+CAML pipelines on the 11-taxon (15\% AD), Mammalian-1X (30\% AD), 10-taxon MH ILS (40\% AD), and 15-taxon H ILS (82\% AD) datasets. It was found that on moderately high and high ILS datasets (i.e., 10-taxon MH ILS and 15-taxon H ILS), WSB+WQMC computed more accurate ASTRAL2 species trees than WSB+CAML. Also, it was found that WSB+WQMC computed better ASTRID species trees than WSB+CAML on the 10-taxon dataset but similar species trees on the 15-taxon dataset. However, on low and medium ILS level datasets (i.e., 11-taxon and Mammalian-1X datasets), WSB-CAML outperformed WSB-WQMC in computing more accurate species trees using both ASTRID and ASTRAL2. It was found that the WSB+WQMC pipeline computed similar or better ASTRAL2 species tree than the basic unbinned pipeline for the 11-taxon L ILS and 15-taxon H ILS datasets, whereas it computed better ASTRID species tree for all datasets used in this experiment except the Mammalian-1X dataset.
  
\begin{figure}[p]
\centering
\includegraphics[width=\textwidth]{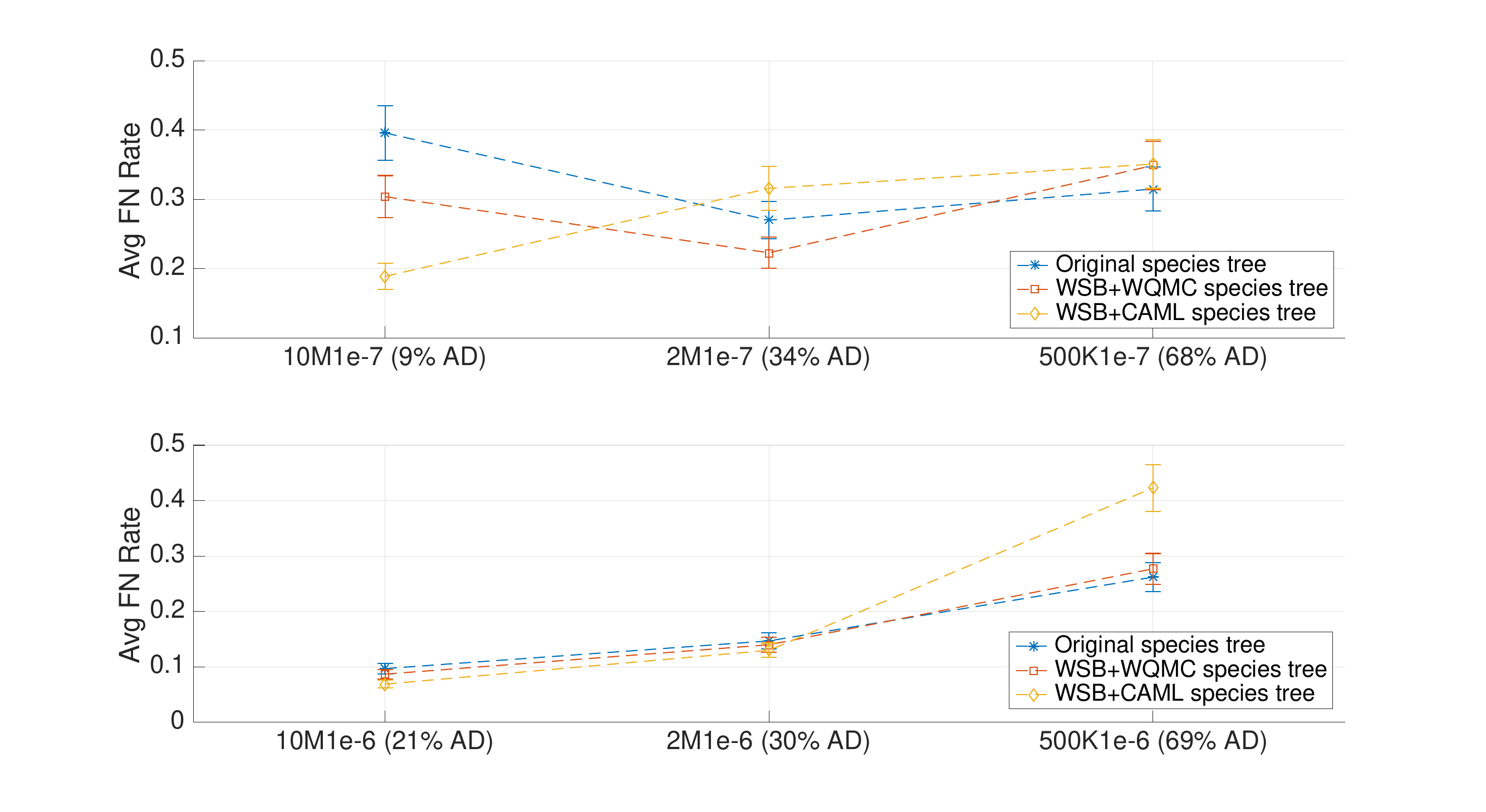}
\caption{\textbf{ASTRID species tree estimation error on SimPhy datasets having speciation rate 1e-7 (top) and having speciation rate 1e-6 (bottom) for running WSB+WQMC and WSB+CAML.} We show results on the 10M1e-7 (9\% AD), 2M1e-7 (34\% AD), 500KM1e-7 (68\% AD), 10M1e-6 (21\% AD), 2M1e-6 (30\% AD), and 500K1e-6 (69\% AD) datasets with 100 genes and 100bp alignment length. WSB+WQMC was run using binning threshold 95\% and confidence value 0.2. WSB+CAML was run using binning threshold 95\%. The accuracy of species tree on WSB+WQMC and WSB+CAML gene trees is also compared to the accuracy of species tree on original gene trees. Average FN rate is shown with standard error bars over 10 replicates.}
\label{fig:caml-simphy-astrid}
\end{figure}

\begin{figure}[p]
\centering
\includegraphics[width=\textwidth]{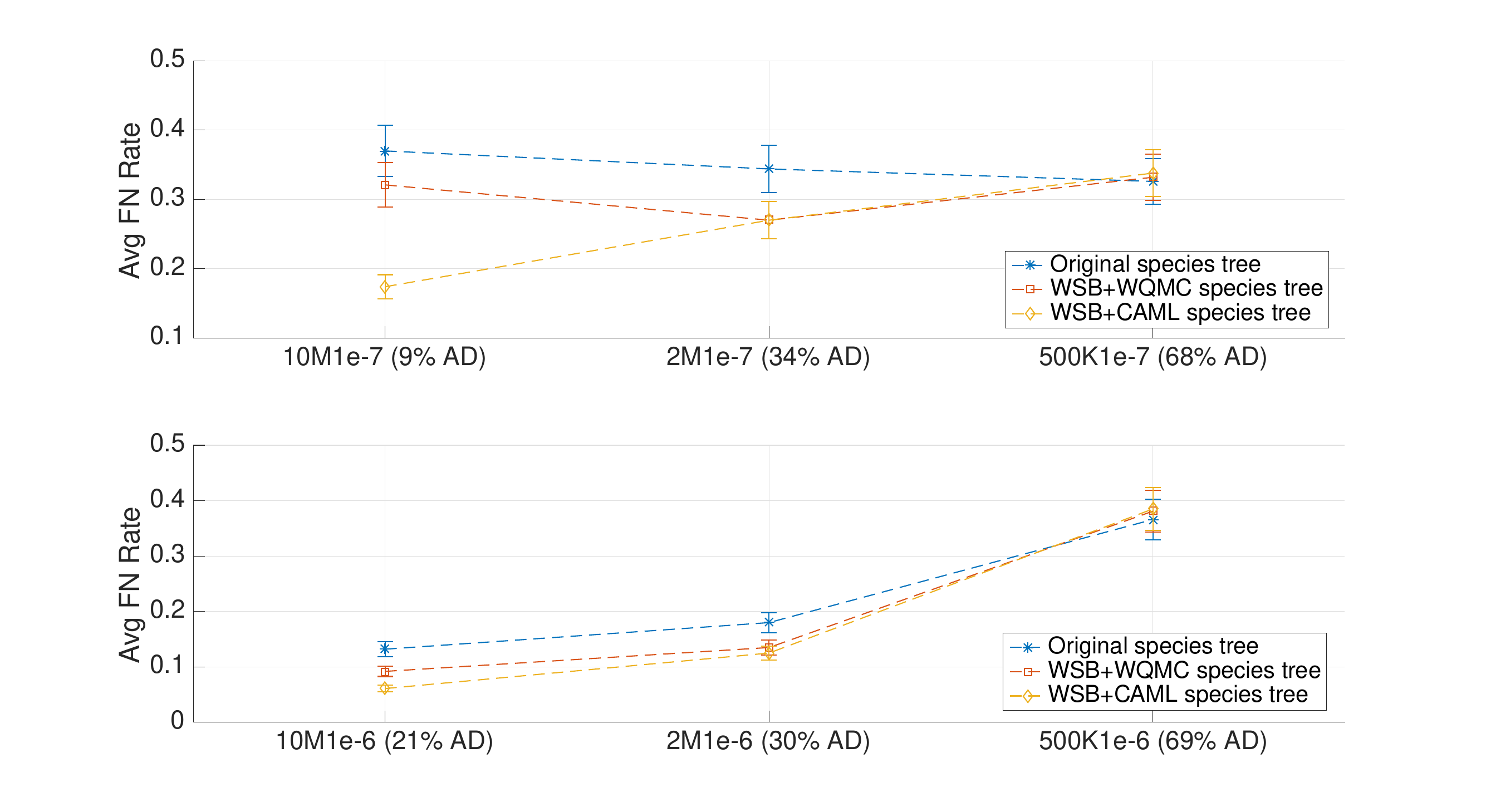}
\caption{\textbf{ASTRAL2 species tree estimation error on SimPhy datasets having speciation rate 1e-7 (top) and having speciation rate 1e-6 (bottom) for running WSB+WQMC and WSB+CAML.} We show results on the 10M1e-7 (9\% AD), 2M1e-7 (34\% AD), 500KM1e-7 (68\% AD), 10M1e-6 (21\% AD), 2M1e-6 (30\% AD), and 500K1e-6 (69\% AD) datasets with 100 genes and 100bp alignment length. WSB+WQMC was run using binning threshold 95\% and confidence value 0.2. WSB+CAML was run using binning threshold 95\%. The accuracy of species tree on WSB+WQMC and WSB+CAML gene trees is also compared to the accuracy of species tree on original gene trees. Average FN rate is shown with standard error bars over 10 replicates.}
\label{fig:caml-simphy-astral2}
\end{figure}

  Figures \ref{fig:caml-simphy-astrid} and \ref{fig:caml-simphy-astral2} show the species tree estimation results by running the WSB+WQMC and WSB+CAML pipelines on the SimPhy datasets having speciation rates 1e-7 and 1e-6. WSB+CAML computed better species trees for low and medium ILS datasets (i.e., 10M1e-7, 10M1e-6, and 2M1e-6) using both ASTRID and ASTRAL2. For moderately high and high ILS datasets (i.e., 2M1e-7, 500K1e-7, and 500K1e-6), WSB+WQMC was found to be at least as accurate as WSB+CAML in computing species trees using both ASTRID and ASTRAL2. Moreover,  the ASTRID trees computed by WSB+WQMC were found to be substantially better than the ASTRID trees computed by WSB+CAML on the SimPhy-2M1e-7 and SimPhy-500K1e-6 datasets. Both pipelines performed better than the basic unbinned pipeline in computing species trees for low and medium ILS levels irrespective of the summary method. In moderately high ILS conditions, WSB+WQMC computed better species trees than the basic unbinned pipeline for both ASTRID and ASTRAL2. However, WSB+CAML computed better ASTRAL2 species trees but worse ASTRID species trees than the basic unbinned pipeline on moderately high ILS datasets. For high ILS datasets, both pipelines computed worse species trees than the species trees on original gene trees for both ASTRID and ASTRAL2. 
  
  In summary, it was found that WSB+WQMC computed similar or better species trees than WSB+CAML in moderately high and high ILS conditions irrespective of the summary method. However, in low and medium ILS conditions, WSB+CAML outperformed WSB+WQMC in computing the species trees using both ASTRID and ASTRAL2.
\newpage
\section{Figures and Tables}
\vfill
\vfill

\begin{figure}[p]
\centering
\includegraphics[width=\textwidth]{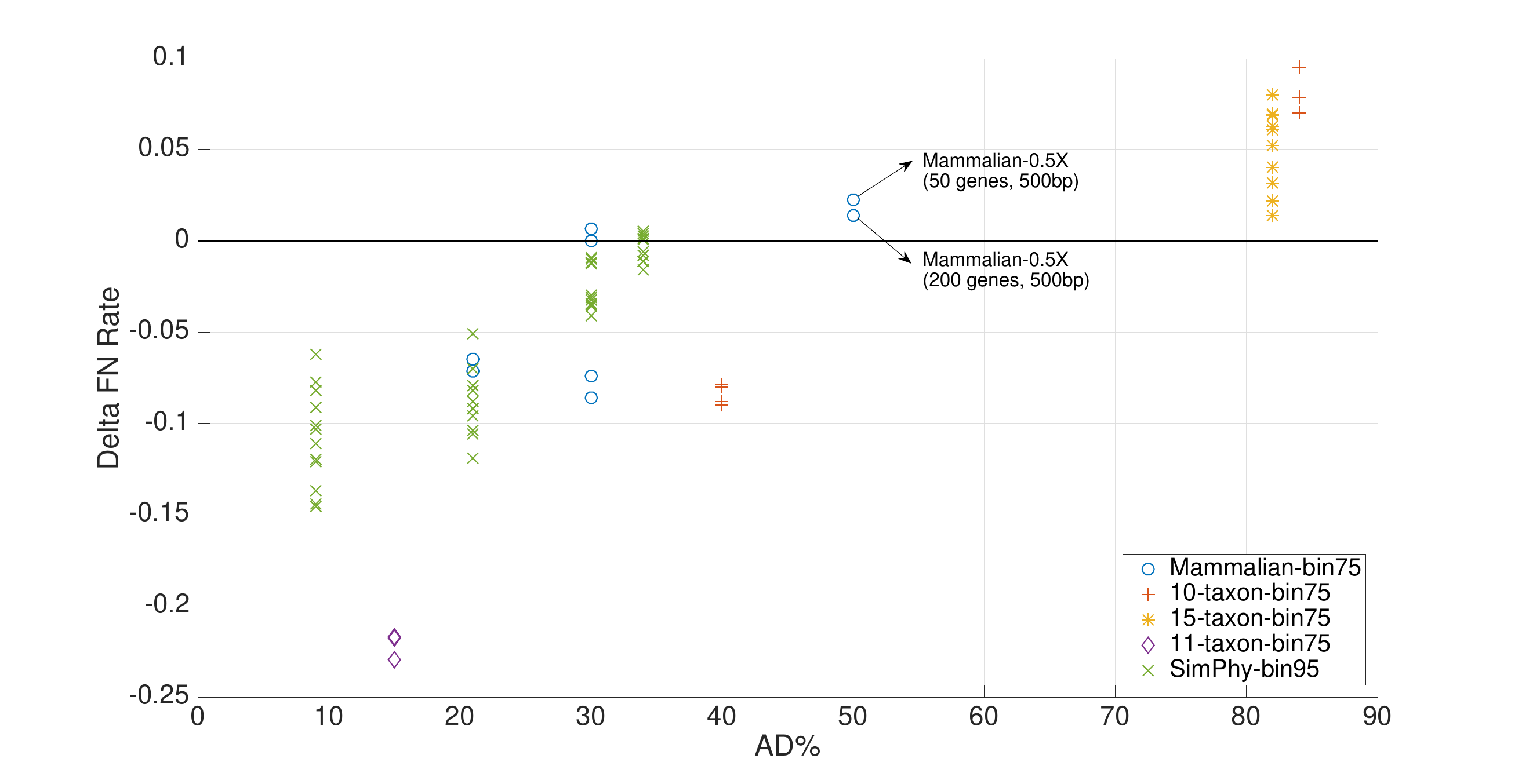}
\caption{\textbf{Gene tree estimation error results on running WSB+WQMC on different model conditions of datasets shown in Tables \ref{table:training} and \ref{table:testing}}. We show results on the Mammalian-0.5X (50\% AD), Mammalian-1X (30\% AD), Mammalian-2X (21\% AD), SimPhy-10M1e-7 (9\% AD), SimPhy-10M1e-6 (21\% AD), SimPhy-2M1e-6 (30\% AD), SimPhy-2M1e-7 (34\% AD), 10-taxon MH ILS (40\% AD), 10-taxon H ILS (84\% AD), 15-taxon H ILS (82\% AD), and 11-taxon L ILS (15\% AD) datasets with different number of genes and alignment lengths, as shown in Tables \ref{table:training} and \ref{table:testing}. The ILS level is varied on the x-axis with increasing ILS from left to right. Delta FN rate was computed by subtracting average FN rate of original gene trees from average FN rate of WSB+WQMC gene trees. Average FN rates were computed over 10 replicates for each dataset and model condition. Each point in the plot corresponds to delta FN rate on some model condition  of the corresponding dataset. WSB+WQMC was run on these datasets with confidence value 0.2 and binning thresholds indicated in the legend.}
\label{fig:testing-gene}
\end{figure}

\clearpage

\clearpage

\begin{figure}[p]
\centering
\includegraphics[width=\textwidth]{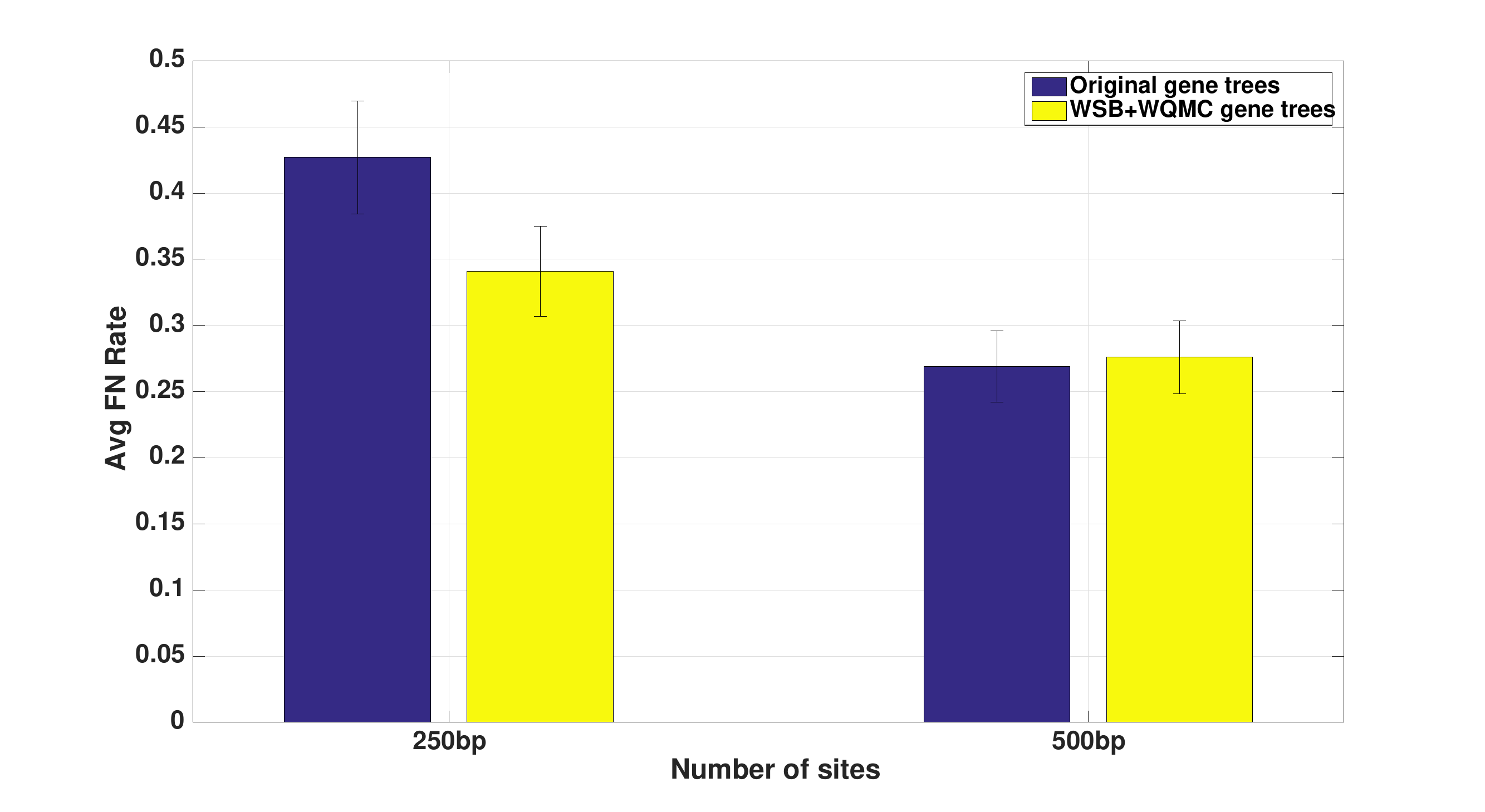}
\caption{\textbf{Gene tree estimation error on the Mammalian-1X dataset with two different alignment lengths for running WSB+WQMC.} We show results on the Mammalian-1X (30\% AD) dataset was used having 200 genes and two different alignment lengths (250bp and 500bp). WSB+WQMC was run using binning threshold 75\% and confidence value 0.2. The accuracy of WSB+WQMC gene trees is also compared to the accuracy of original gene trees. Average FN rate is shown with standard error bars over 10 replicates.}
\label{fig:training-numsites-gene}
\end{figure}

\begin{figure}[p]
\centering
\includegraphics[width=\textwidth]{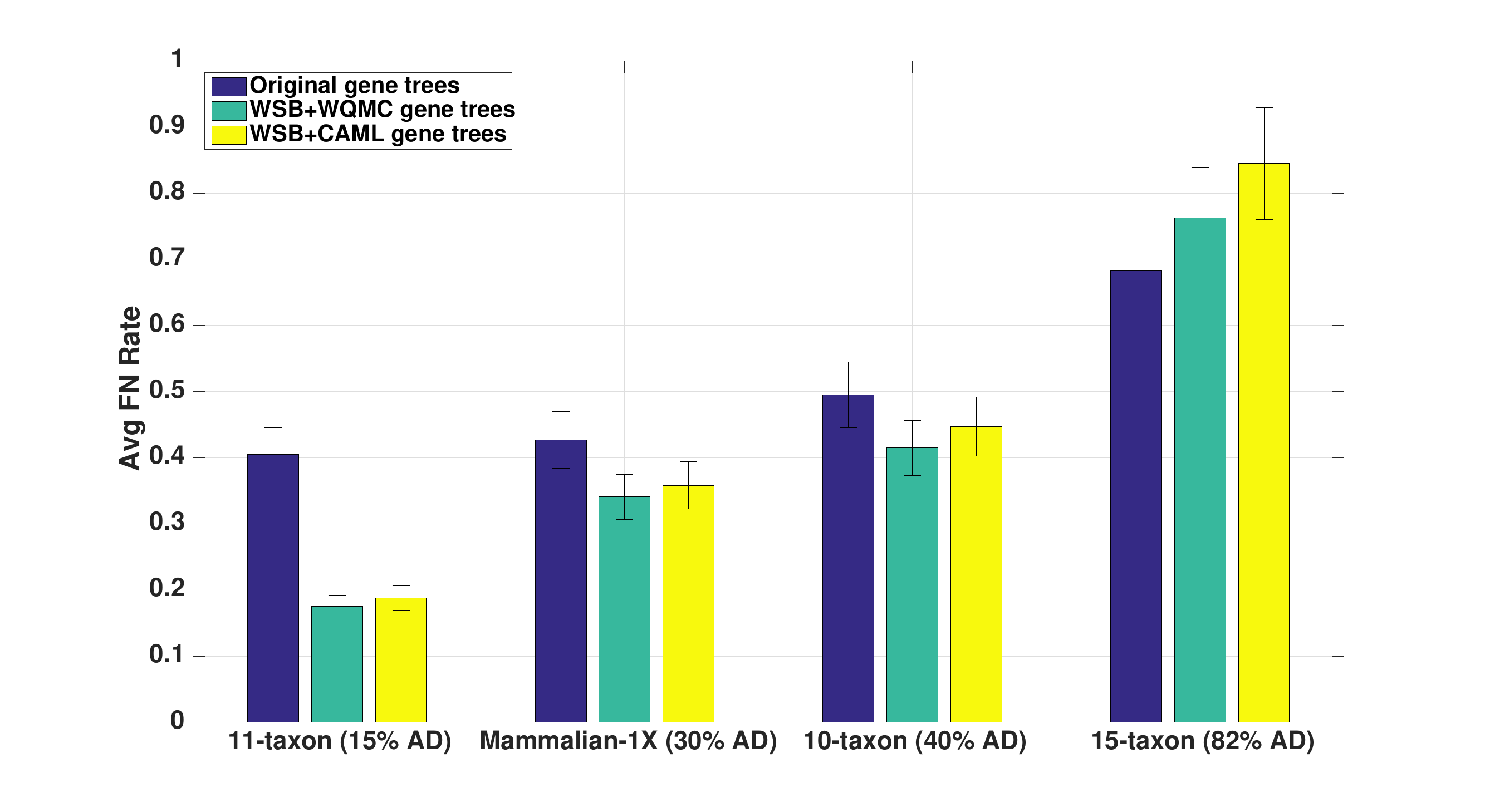}
\caption{\textbf{Gene tree estimation error on 11-taxon, Mammalian-1X, 10-taxon and 15-taxon datasets for running WSB+WQMC and WSB+CAML.} We show results on the 11-taxon L ILS (100 genes and 500bp),  Mammalian-1X (200 genes, 250bp), 10-taxon MH ILS (100 genes, 100bp), and 15-taxon H ILS (100 genes and 100bp) datasets were used. The level of ILS increases from left to right. WSB+WQMC was run using binning threshold 75\% and confidence value 0.2. WSB+CAML was run using binning threshold 75\%. The accuracy of WSB+WQMC and WSB+CAML gene trees were also compared to the accuracy of original gene trees. Average FN rate is shown with standard error bars over 10 replicates.}
\label{fig:caml-others-gene}
\end{figure}

\begin{figure}[p]
\centering
\includegraphics[width=\textwidth]{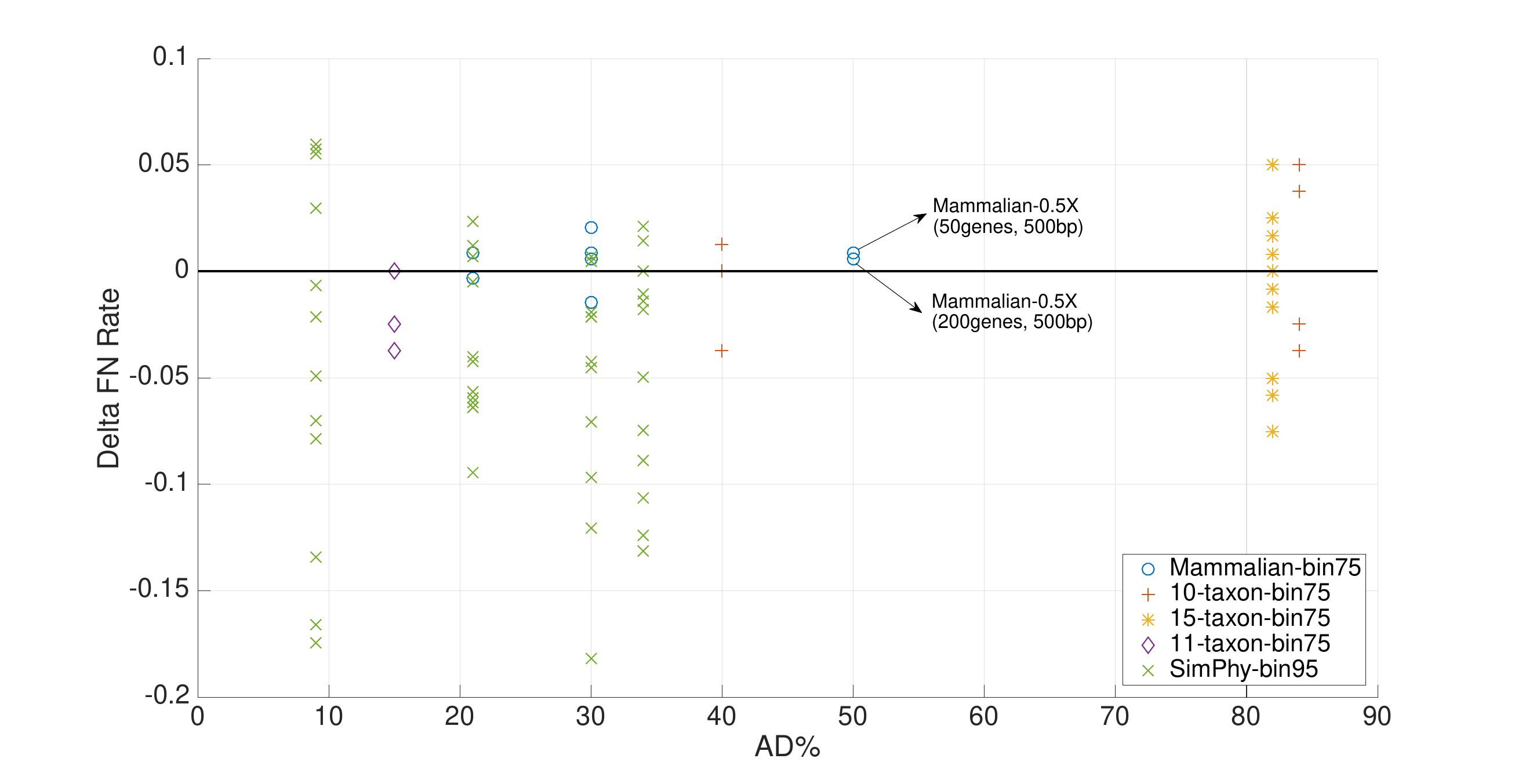}
\caption{\textbf{ASTRAL2 species tree estimation error results on running WSB+WQMC on different model conditions of datasets shown in Tables \ref{table:training} and \ref{table:testing}}. We show results on the Mammalian-0.5X (50\% AD), Mammalian-1X (30\% AD), Mammalian-2X (21\% AD), SimPhy-10M1e-7 (9\% AD), SimPhy-10M1e-6 (21\% AD), SimPhy-2M1e-6 (30\% AD), SimPhy-2M1e-7 (34\% AD), 10-taxon MH ILS (40\% AD), 10-taxon H ILS (84\% AD), 15-taxon H ILS (82\% AD), and 11-taxon L ILS (15\% AD) datasets with different number of genes and alignment lengths, as shown in Tables \ref{table:training} and \ref{table:testing}. ILS level is varied on the x-axis with increasing ILS from left to right. Delta FN rate was computed by subtracting average FN rate of species tree on original gene trees from average FN rate of species tree on WSB+WQMC gene trees. Average FN rates were computed over 10 replicates for each dataset and model condition. Each point in the plot corresponds to delta FN rate on some model condition  of the corresponding dataset. WSB+WQMC was run on these datasets with confidence value 0.2 and binning thresholds indicated in the legend.}
\label{fig:testing-astral2}
\end{figure}

\begin{figure}[p]
\centering
\includegraphics[width=\textwidth]{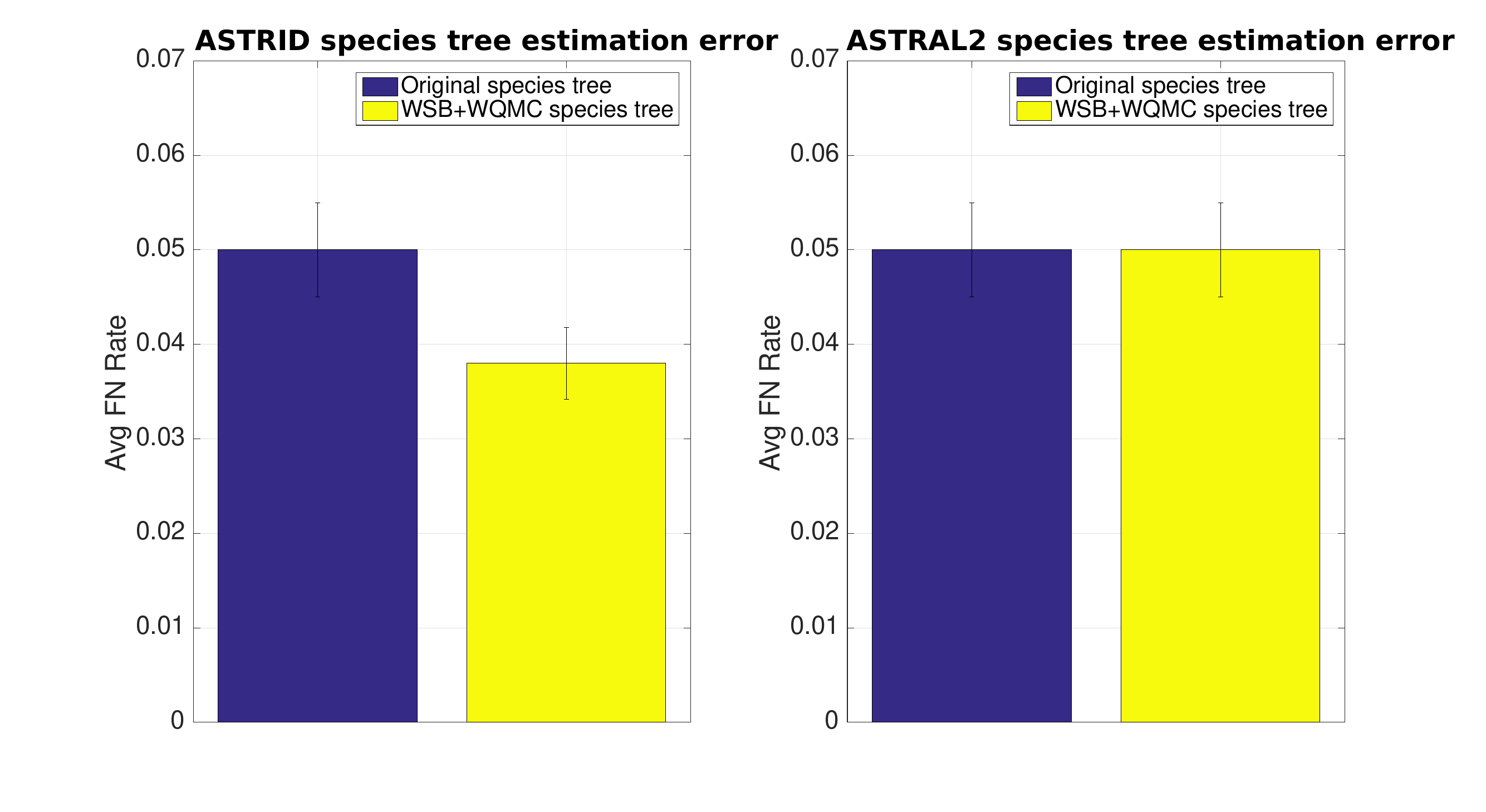}
\caption{\textbf{Species tree estimation error using ASTRID (left) and ASTRAL2 (right) on 11-taxon dataset for running WSB+WQMC.} We show results on the 11-taxon L ILS (15\% AD) dataset was used having 100 genes and 500bp alignment length. WSB+WQMC was run using binning threshold 75\% and confidence value 0.2. The accuracy of WSB+WQMC gene trees and species tree is also compared to the accuracy of original gene trees and species tree. Average FN rate is shown with standard error bars over 10 replicates.}
\label{fig:testing-11taxon-species}
\end{figure}

\begin{figure}[p]
\centering
\includegraphics[width=\textwidth]{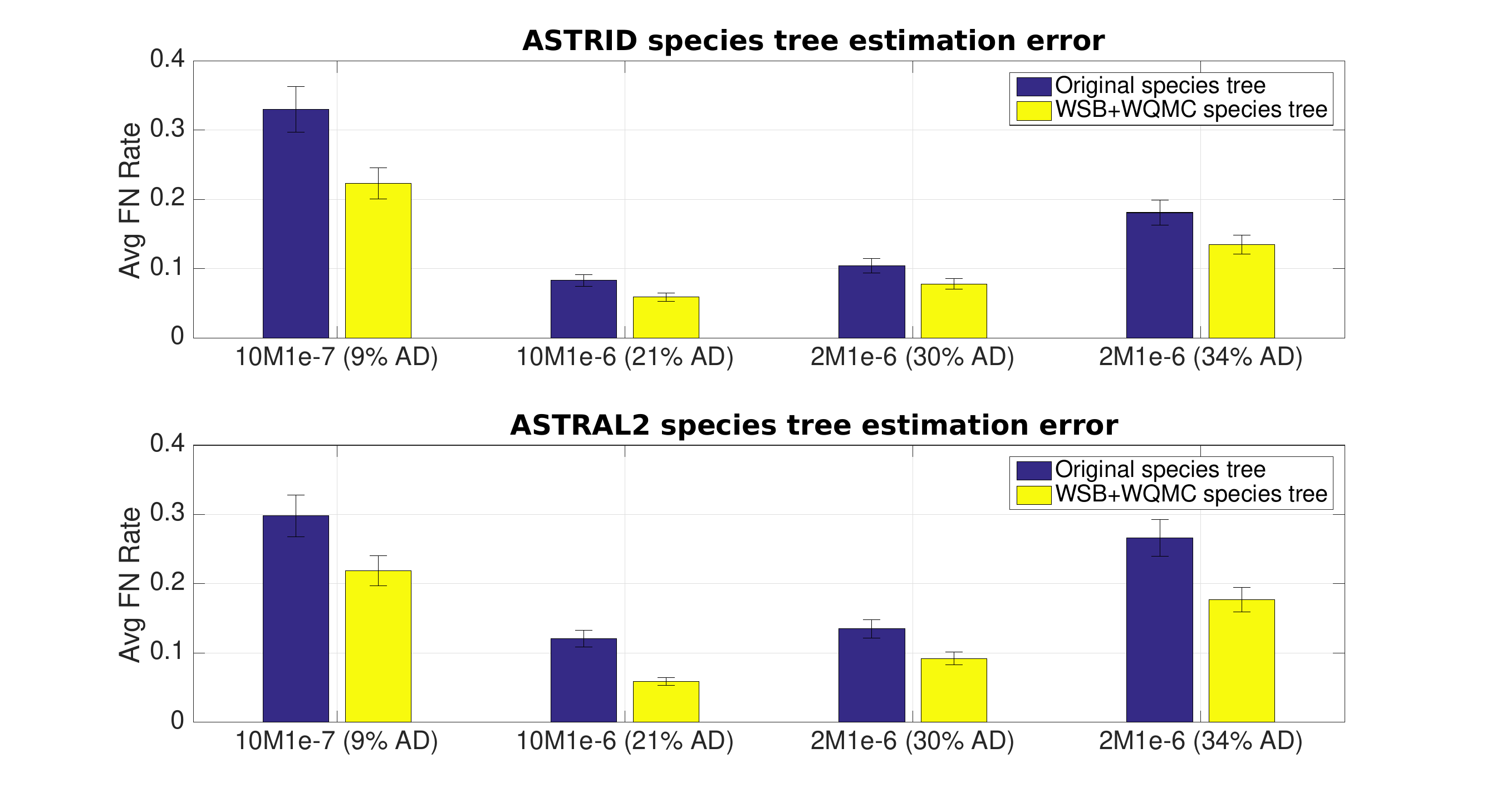}
\caption{\textbf{Species tree estimation error using ASTRID (top) and ASTRAL2 (bottom) on SimPhy datasets for running WSB+WQMC.} We show results on the 10M1e-7 (9\% AD), 10M1e-6 (21\% AD), 2M1e-6 (30\% AD), and 2M1e-7 (34\% AD)  datasets with 200 genes and 200bp alignment length. WSB+WQMC was run using binning threshold 95\% and confidence value 0.2. The accuracy of species tree on WSB+WQMC gene trees is also compared to the accuracy of species tree on original gene trees. Average FN rate is shown with standard error bars over 10 replicates.}
\label{fig:testing-simphy-species}
\end{figure}

\begin{figure}[p]
\centering
\includegraphics[width=\textwidth]{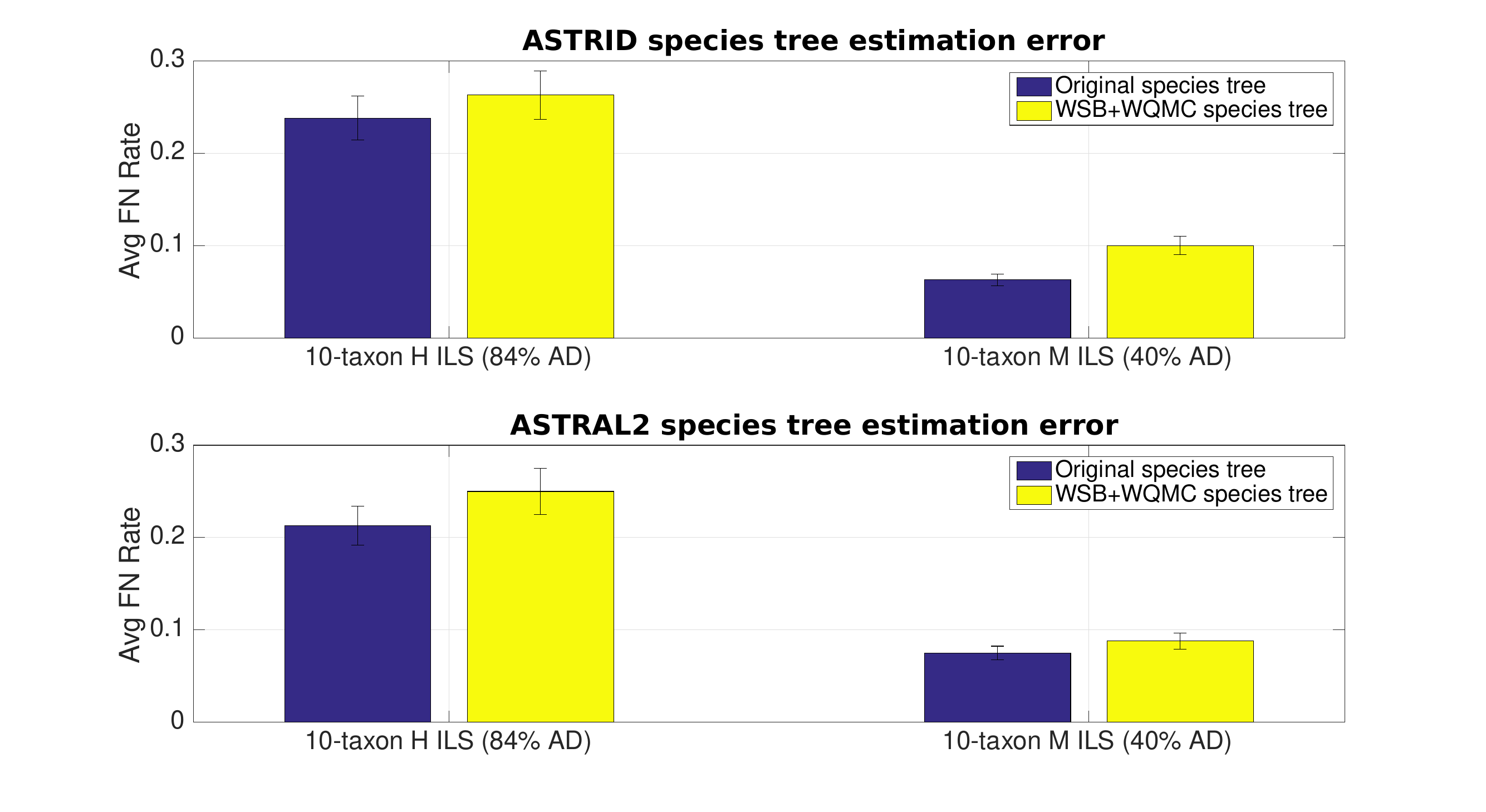}
\caption{\textbf{Species tree estimation error using ASTRID (top) and ASTRAL2 (bottom) on 10-taxon datasets for running WSB+WQMC.} We show results on the 10-taxon MH ILS (40\% AD) and 10-taxon H ILS (84\% AD) with 200 genes and 100bp alignment length. WSB+WQMC was run using binning threshold 75\% and confidence value 0.2. The accuracy of species tree on WSB+WQMC gene trees is also compared to the accuracy of species tree on original gene trees. Average FN rate is shown with standard error bars over 10 replicates.}
\label{fig:testing-10taxon-species}
\end{figure}

\begin{figure}[p]
\centering
\includegraphics[width=\textwidth]{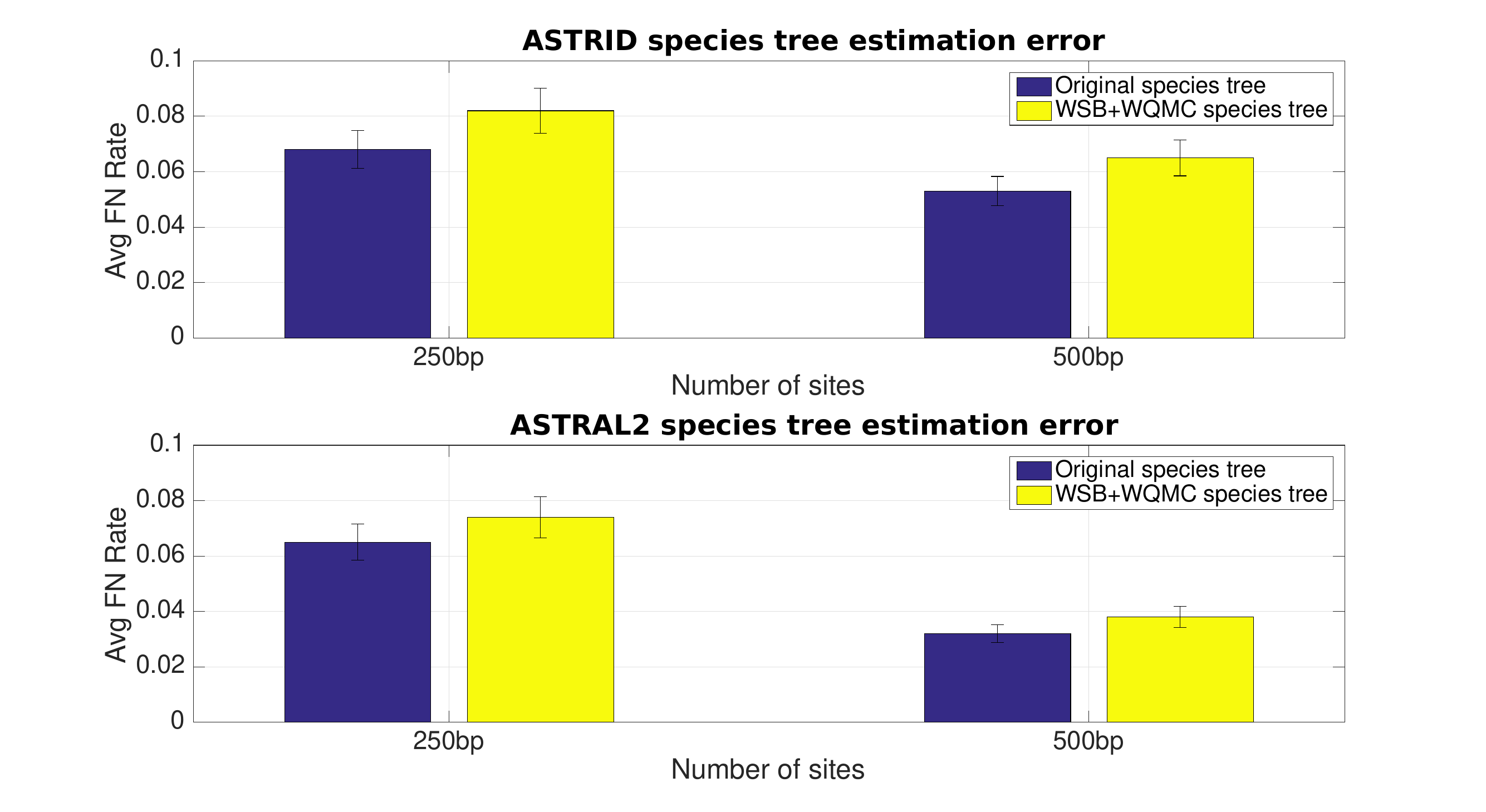}
\caption{\textbf{Species tree estimation error using ASTRID (top) and ASTRAL2 (bottom) on the Mammalian-1X dataset with two different alignment lengths for running WSB+WQMC.} We show results on the Mammalian-1X (30\% AD) dataset was used having 200 genes and two different alignment lengths (250bp and 500bp). WSB+WQMC was run using binning threshold 75\% and confidence value 0.2. The accuracy of species tree on WSB+WQMC gene trees is also compared to the accuracy species tree on original gene trees. Average FN rate is shown with standard error bars over 10 replicates. }
\label{fig:training-numsites-species}
\end{figure}

\end{doublespace}
\chapter{Discussion}
\begin{doublespace}
In this thesis, the WSB+WQMC pipeline was proposed and evaluated on many simulated datasets using BestML analysis. The WSB+CAML pipeline proposed in \cite{bayzid2015weighted} was also evaluated on those datasets and compared with the WSB+WQMC pipeline using BestML analysis. To the best of my knowledge, this is the first study to evaluate the effect of phylogenetic pipelines that use any variants of the statistical binning technique in the BestML paradigm. The prior studies of statistical binning \cite{mirarab2014statistical} and weighted statistical binning \cite{bayzid2015weighted} only evaluated these pipelines using MLBS, which may not be the best way to compute gene trees and species tree in certain model conditions. In \cite{mirarab2014evaluating}, it was seen that BestML computed more accurate species trees than MLBS in model conditions having large enough number of genes. This clearly highlights the importance of the experimental study conducted in this paper, which provides performance results of these binning-based pipelines in BestML paradigm for the first time.   
\section{Summary of observations}
We begin with a comparison of the WSB+WQMC pipeline to the basic unbinned pipeline. In general, it was found that WSB+WQMC improved the accuracy of gene trees (see Figure \ref{fig:testing-gene}) in all model conditions of datasets having low ILS and in most of the model conditions of datasets having medium and moderately high levels of ILS. On the other hand, in high ILS conditions WSB+WQMC computed gene trees that were worse than original gene trees. WSB+WQMC also improved the accuracy of species tree estimation using both ASTRID and ASTRAL2 on most model conditions of datasets having low, medium, and moderately high ILS conditions (see Figures \ref{fig:testing-astral2}, \ref{fig:testing-11taxon-species}, and \ref{fig:testing-simphy-species}), with a few exceptions. Similar to gene tree estimation, it was observed that on most model conditions of datasets having high levels of ILS, WSB+WQMC computed worse species trees than the species trees on original gene trees for both ASTRAL2 and ASTRID (see Figures \ref{fig:testing-astral2} and  \ref{fig:testing-10taxon-species}). 

A comparison of WSB+WQMC to WSB+CAML in terms of gene tree estimation error provides some interesting differences. In high ILS datasets, WSB+WQMC was found to compute more accurate gene trees than WSB+CAML. In moderately high ILS conditions WSB+WQMC was found to be at least as good as WSB+CAML. In low and medium ILS datasets, there was no clear winner and the relative performance of WSB+CAML and WSB+WQMC depends on the dataset itself.

When comparing WSB+WQMC and WSB+CAML in terms of species tree estimation more interesting trends were observed. In high ILS datasets, WSB+WQMC was at least as accurate as  WSB+CAML for computing species trees using both ASTRID and ASTRAL2. In moderately high ILS conditions, again WSB+WQMC was found to be at least as good as WSB+CAML, irrespective of the summary method. In low and medium ILS datasets WSB+CAML computed better species trees than WSB+WQMC using both ASTRAL2 and ASTRID.

% The experiments in the paper further studied the effects of binning threshold, ILS level and alignment length on accuracy of WSB+WQMC and observed similar trends as reported by authors of Naive binning \cite{bayzid2013naive}, Statistical Binning \cite{mirarab2014statistical} and Weighted statistical binning \cite{bayzid2015weighted}.

\section{Impact of ILS level on WSB+WQMC and WSB+CAML}
For low enough ILS levels, both WSB+WQMC gene trees and WSB+CAML gene trees were more accurate than original gene trees, but as the ILS level increased the improvement decreased, until finally the re-estimated gene trees were less accurate. For species tree estimation, a similar relationship with ILS level was observed. This relationship of accuracy of binning based pipelines and ILS level was observed in previous studies as well. In \cite{bayzid2013naive}, the naive binning pipeline was shown to improve species tree estimation in low ILS conditions. Similarly, both statistical binning and weighted statistical binning improved species tree estimation over the basic unbinned pipeline for low ILS conditions. 

This relationship of ILS level and performance of WSB+WQMC and WSB+CAML is expected, as in low ILS conditions the bins computed by both pipelines will have genes having very similar evolutionary histories, and thus, using data from genes within the same bin will boost the phylogenetic signal and won't lead to any problematic analysis. However, when the level of ILS is high and sequences aren't very long, the binning step in these pipelines will group genes from different evolutionary histories into the same bin, leading to poor performance of both WSB+CAML and WSB+WQMC. This will happen mainly because many conflicting bipartitions between genes will not be detected in the binning step due to low bootstrap support values of edges in the initially estimated gene tree. 

\section{Impact of alignment length on WSB+WQMC}
In this thesis it was observed that the magnitude of improvement in gene tree and species tree estimation when comparing the WSB+WQMC pipeline to the basic unbinned pipeline was greater for shorter alignment lengths. This trend was also seen in \cite{bayzid2015weighted}, where it was seen that the difference between species tree estimation error by running the WSB+CAML pipeline and the basic unbinned pipeline decreased with increasing sequence length. This trend makes sense as when the alignment lengths are longer bootstrap support values of edges in the estimated gene trees increase, leading to lower bin sizes for the same binning threshold. Also, as the alignment becomes longer, the initially estimated gene tree improves and amount of improvement that can be achieved using binning pipelines  is low. 

\section{Impact of binning threshold on WSB+WQMC}
In this thesis, it was observed that binning 75\% outperformed binning 100\% in medium and high ILS conditions, with binning 100\% being much worse and computing highly inaccurate gene trees and ASTRAL2 species tree for high ILS dataset. This is mainly because having a very high binning threshold (putting all genes into a single bin in this case) can put  genes with very different evolutionary histories into the same bin. On the other hand, it was also seen that binning 100\% performed better than binning 75\% for low ILS datasets. This result can be attributed to the fact that lower binning threshold reduces bin sizes, which results in less data being used for each gene to compute a new gene tree. The limited amount of data used particularly hurts in low ILS conditions where all gene trees have very similar histories and having larger bin sizes can substantially improve performance of binning-based pipelines.

\section{Relationship between gene tree and species tree estimation error}
As expected, for most model conditions studied in this paper, improving gene trees using both WSB+WQMC and WSB+CAML leads to improvement in species tree estimation. However, it was observed that for some cases improving gene trees using WSB+WQMC and WSB+CAML didn't improve ASTRID and ASTRAL2 species trees when compared to the species trees on original gene trees. For example, the WSB+WQMC pipeline on the Mammalian-1X dataset computed better gene trees but worse species trees (see Figures \ref{fig:training-numsites-gene} and \ref{fig:training-numsites-species}). Moreover, WSB+CAML improved gene tree estimation on the 10-taxon dataset, but made species trees worse (see Figures \ref{fig:caml-others-gene}, \ref{fig:caml-others-astrid} and \ref{fig:caml-others-astral2}).

This was an unexpected result and conflicted with the common belief that reducing gene tree estimation error leads to reduction in species tree estimation error. The reason for this behavior remains unclear and needs further investigation in  the future.

Another anomaly was observed when both WSB+WQMC and WSB+CAML improved ASTRAL2 and ASTRID species trees for the 15-taxon H ILS dataset even after increasing gene tree estimation error (see Figures \ref{fig:caml-others-gene}, \ref{fig:caml-others-astrid} and \ref{fig:caml-others-astral2}). 15-taxon H ILS datasets have strict molecular clock and it is suspected that it can be leading to this strange result.

\section{MLBS vs. BestML}
The impact of ILS level and alignment length on the performance of binning-based pipelines were similar for both BestML and MLBS analyses. However, we found some important differences between BestML and MLBS that need to be discussed. In \cite{bayzid2015weighted}, it was found that the statistical binning and WSB+CAML pipelines decreased species tree accuracy compared to the basic unbinned pipeline only in a very few model conditions having ultra high levels of ILS (more than 80\% AD) and small numbers of taxa. On the other hand, in BestML analyses we found that both WSB+WQMC and WSB+CAML were found to increase species tree estimation error compared to the basic unbinned pipeline more frequently and even in model conditions having 40\% AD ILS level (see Figures \ref{fig:caml-others-astrid} and \ref{fig:caml-others-astral2}). The benefit of using BestML over MLBS or vice versa is not clear and requires further research.

\section{Using FastTree support values instead of bootstrap support values}

The original gene trees on the SimPhy datasets were estimated without bootstrap support values. In order to run the WSB+WQMC and WSB+CAML pipelines on these datasets, FastTree local support values instead of bootstrap support values were used. FastTree support values tend to be much higher than bootstrap support values, which resulted in a very high bootstrap threshold value of 95\% being used for the SimPhy datasets. Even though FastTree local support values of an edge gives some indication of the reliability of that edge, they are very different from bootstrap support values and are suspected to be of poor quality. Due to the use of FastTree local support values instead of bootstrap support values, there is a possibility that the trends seen on the SimPhy datasets may not hold if  bootstrap support values are used in the future analysis.

\section{Limitations of WSB+WQMC}
Even though the WSB+WQMC pipeline decreased gene tree  and species tree estimation error on many datasets studied in this paper, it has a few limitations. Similar to the WSB+CAML pipeline, it performed poorly in high ILS conditions. Though it performed better than WSB+CAML in all high ILS datasets we examined, many times it was better to not use WSB+WQMC compared to the basic unbinned pipeline. This poor performance in high ILS conditions can be attributed to putting highly varied genes into a single bin, a common problem haunting binning-based pipelines. 

Another drawback of WSB+WQMC is its need for the confidence value $c$ and the binning threshold $t$. These values have a significant impact on the performance of WSB+WQMC, but we do not yet have a way of determining the optimal values that should be used in an experiment. In this study, the training phase was used to narrow down confidence values based on  a  subset of data and  that value was used for the rest of the data. Even though we were able to obtain a good confidence value to be used for all the datasets in the study, the optimal confidence value can be  different for different datasets.

\end{doublespace}
\chapter{Summary}
\begin{doublespace}
Gene tree estimation and species tree estimation are known to suffer when individual loci have low phylogenetic signal. In most phylogenetic studies, the limited amount of data causes the gene alignments to have low phylogenetic signal, which in turn leads to poorly estimated gene trees.  Because species trees and gene trees can differ due to incomplete lineage sorting, the estimation of species trees requires multiple loci. In the presence of ILS, many summary methods like ASTRAL2, MP-EST, and ASTRID can be used to estimate the species tree by combining gene trees. However, this study, as well as others \cite{mirarab2014astral, leache2010accuracy}, suggests  that  species trees computed by summary methods using poorly estimated gene trees have low accuracy.  Thus, an attempt to improve gene trees by enhancing phylogenetic signal helps in both gene tree and species tree estimation.

Three methods (naive binning, statistical binning, and weighted statistical binning) have been developed to improve gene tree and species tree estimation when individual gene alignments have low phylogenetic signal. The statistical binning pipeline proposed in \cite{mirarab2014statistical} was found to improve species tree estimation using MLBS (multi-locus bootstrapping) analysis. However, in  \cite{avni2015weighted}, the statistical binning pipeline was shown to be statistically inconsistent under the GTR+MSC model. The weighted statistical pipeline \cite{bayzid2015weighted} improved on the statistical binning pipeline and was shown to be statistically consistent under the GTR+MSC model. In \cite{bayzid2015weighted}, the weighted statistical binning pipeline was evaluated on many datasets using MLBS analysis. When compared to the basic unbinned pipeline, the weighted statistical binning pipeline (referred as WSB+CAML in this thesis) was found to compute more accurate ASTRAL2 and MP-EST species trees except for a few datasets having extremely high ILS and small numbers of taxa.

Prior to this study, the weighted statistical binning pipeline was only tested using MLBS analysis, leaving it untested when it was run using BestML analysis. Although MLBS is more frequently used for species tree estimation, it may not be the most accurate way to compute species trees. The results in \cite{mirarab2014evaluating} showed that BestML computed more accurate species trees than MLBS with different species tree estimation methods on datasets having large enough number of genes. This indicates the importance of evaluating weighted statistical binning and other phylogenetic pipelines using BestML analysis. 

In this thesis, a novel phylogenetic pipeline named WSB+WQMC to improve gene tree and species tree estimation in the presence of low phylogenetic signal is proposed. The WSB+WQMC pipeline shares several design features with the WSB+CAML pipeline, but has some other desirable properties. The WSB+WQMC pipeline has good theoretical guarantees and is shown to be statistically consistent under the GTR+MSC model when a slightly different version of WQMC \cite{avni2015weighted} is used in the pipeline. 

An experimental study was conducted in this thesis to evaluate the WSB+WQMC pipeline. It also compared WSB+WQMC to the WSB+CAML pipeline in terms of gene tree and species tree estimation on various datasets. This study was the first to evaluate the weighted statistical binning pipeline using BestML analysis and found that both WSB+WQMC and WSB+CAML followed trends seen in
MLBS analysis with some important differences. 

When compared to the initially estimated gene trees, both WSB+WQMC and WSB+CAML pipelines were found to help in improving gene tree estimation in low ILS conditions. However, the level of ILS had a big effect on the performance of both pipelines, and it was found that both WSB+WQMC and WSB+CAML can make gene trees worse on high ILS datasets.

Furthermore, when compared to ASTRID and ASTRAL2 species trees on original gene trees, both WSB+WQMC  and WSB+CAML pipelines again improved species tree estimation on most of the datasets studied. But, on a few datasets, both pipelines ended up reducing the accuracy of species trees. This result was unexpected and refutes the common belief that reducing gene tree estimation error always reduces species tree estimation error. Therefore, care must be taken when using both WSB+WQMC and WSB+CAML pipelines as they may not always turn out to be beneficial. 

When comparing WSB+WQMC and WSB+CAML in terms of gene tree and species tree estimation, no clear winner was found. The relative performance of WSB+WQMC and WSB+CAML was dependent on the dataset and the level of ILS. In low and medium levels of ILS, it was found that WSB+CAML computed better species trees than WSB+WQMC, but there was no clear winner for gene tree estimation. On the other hand, in moderately high and high levels of ILS, WSB+WQMC was at least as good as WSB+CAML for both gene tree and species tree estimation. 

This study clearly suggests the potential benefits of using both WSB+CAML and WSB+WQMC pipelines for gene tree and species tree estimation. However, there are a few cases where both these pipelines can rather hurt in estimating gene trees and species tree. This, warrants the need for refinements in both WSB+WQMC and WSB+CAML pipelines along with research in the direction of using binning-based pipelines to boost phylogenetic signal.

This thesis highlights various methods to improve gene tree and species tree estimation in the presence of low phylogenetic signal. However, the decision to use binning-based pipelines rather than the basic unbinned pipeline is not straightforward and can have consequences. Moreover, the choice of using MLBS or BestML for running these pipelines is also not clear and requires further research.  

One major challenge in this study was evaluating the performance of WSB+WQMC and WSB+CAML pipelines on the SimPhy datasets. The original gene trees for these datasets did not have bootstrap support values on edges and FastTree support values were used instead in the analysis. FastTree support values are suspected to be a lower quality estimator for edge reliability than bootstrap support values. This casts some doubts on the results obtained on the SimPhy datasets and demonstrates the limit of using these datasets in this thesis.

The WSB+WQMC pipeline is not perfect and has scope for improvements. It requires the confidence value and the binning threshold value to be set. Research in the direction of automatically inferring these parameters may substantially improve the accuracy of the pipeline by being able to pick optimal values for these parameters based on the data. The WSB+WQMC pipeline used WQMC as the heuristic to estimate gene trees from the set of weighted quartets and it would be interesting to explore the performance of the pipeline using other heuristics. The WSB+WQMC pipeline performed poorly in most high ILS datasets; thus, WSB+WQMC needs to be further improved in the future to avoid decreasing accuracy of gene trees in high ILS conditions. 

This study only investigated the performance of WSB+WQMC on simulated datasets. The performance of the WSB+WQMC pipeline on biological datasets needs to be evaluated in the future. It was seen that binning threshold value has an impact on binning pipelines and research aiming at determining good threshold values would be beneficial. One question that isn't addressed in this study is whether it is better to use BestML or MLBS analysis for binning-based pipelines. Evaluating the WSB+WQMC and WSB+CAML pipelines using MLBS analysis can help answer this question. Finally, since concatenation is known to be computationally expensive, it is conjectured that WSB+WQMC is faster and more scalable than WSB+CAML. Analysis of running time and memory usage of the WSB+CAML and WSB+WQMC pipelines can help confirm our suspicion and needs to be done in the future.
\end{doublespace}
\bibliographystyle{unsrt}
\bibliography{references}
%\chapter*{Appendix}
%\addcontentsline{toc}{chapter}{Appendix}
\end{document}